\documentclass[%
 preprint, 
 amsmath,amssymb,
 aps, physrev,
pra,
]{revtex4-2}

\usepackage{graphicx}
\usepackage{dcolumn}
\usepackage{bm}
\usepackage{amsmath}       
\usepackage{algorithm}     
\usepackage{algorithmic}   
\usepackage{amssymb}       
\usepackage{amsthm}        
\usepackage{float}
\usepackage{setspace}
\usepackage{mathrsfs}
\usepackage{booktabs}   
\usepackage{multirow}    
\usepackage{tabularx}  
\usepackage{booktabs}   
\usepackage{multirow}   
\usepackage{array}      
\usepackage{booktabs}
\usepackage{dcolumn}
\usepackage{amsmath}
\usepackage{xcolor}

\usepackage{hyperref}

\begin{document}

\preprint{APS/123-QED}

\title{\textbf{Channel-Constrained Markovian Quantum Diffusion Model from Open System Perspective}}

\author{Qin-Sheng Zhu}
\email{zhuqinsheng@uestc.edu.cn}
\affiliation{%
	School of Physics, University of Electronic Science and Technology of China, Cheng Du, 610054, China
}%
\affiliation{%
	Institute of Electronics and Information Industry Technology of Kash, Kash, 844000, China
}%

\author{Geng Chen}
\email{Corresponding author: 202312081614@std.uestc.edu.cn.
	\\Author Contributions: Q.-S. Zhu and G. Chen contributed equally to this work.}
\affiliation{%
	School of Computer Science and Engineering, University of Electronic Science and Technology of China, Cheng Du, 610054, China
}%

\author{Lian-Hui Yu}
\affiliation{%
	School of Computer Science and Engineering, University of Electronic Science and Technology of China, Cheng Du, 610054, China
}%

\author{Xiaodong Xing}
\affiliation{School of Quantum Information Future Technology, Henan University, Zhengzhou 450046, China and Henan Key Laboratory of Quantum Materials and Quantum Energy, Henan University, Zhengzhou 450046, China and Institute of Quantum Materials and Physics, Henan Academy of Sciences, Zhengzhou, 450046 China}

\author{Xiao-Yu Li}
\affiliation{%
	School of Information and Software Engineering, University of Electronic Science and Technology of China, Cheng Du, 610054, China
}%
\affiliation{%
	Institute of Electronics and Information Industry Technology of Kash, Kash, 844000, China
}%

\date{\today}

\begin{abstract}

We present a channel-constrained Markovian quantum diffusion (CCMQD) model that prepares quantum states by rigorously framing the generative process within the dynamics of open quantum systems. Our model interprets the forward diffusion process as natural decoherence using quantum master equations, whereas the reverse denoising is achieved by learning inverse quantum channels. Our core innovation is a comprehensive channel-constrained framework: we model the diffusion and denoising steps as quantum channels defined by Kraus operators, ensure their physical validity through optimization on the Stiefel manifold, and introduce tailored training strategies and loss functions that leverage this constrained structure for high-fidelity state reconstruction. Experimental validation on systems ranging from single qubits to entangled states $7$ -qubits demonstrates high-fidelity state generation, achieving fidelities exceeding $0.998$ under both random and depolarizing noise conditions. This work confirms that quantum diffusion can be characterized as a controlled Markov evolution, demonstrating that environmental interactions are not limited to being a source of decoherence but can also be utilized to achieve high-fidelity quantum state synthesis.
\end{abstract}

\maketitle

\newpage

\section{Introduction}
Quantum generative modeling \cite{tian2023recent} represents an essential approach for preparing complex quantum states and learning quantum distributions, with applications ranging from quantum simulation \cite{zhu2022generative} to state preparation \cite{niu2022entangling}. Inspired by the classical diffusion model \cite{croitoru2023diffusion, chen2023diffusiondet, kingma2021variational}, various quantum generative models have been proposed in recent years, for instance Quantum Generative Adversarial Networks (QGANs) \cite{lloyd2018quantum,dallaire2018quantum,ma2025quantum}, Quantum Circuit Born Machines (QCBMs) \cite{liu2018differentiable, gili2023quantum}, and Quantum Variational Autoencoders (QVAEs) \cite{khoshaman2018quantum,rocchetto2018learning}. The essential logic of these generative models is to learn the underlying probability distribution from a given data, and generating new data that follows the distribution. These  models have demonstrated remarkable performance in generating quantum states \cite{dallaire2018quantum, chakrabarti2019quantum, braccia2021enhance,niu2022entangling}, and learning complex quantum distributions \cite{furrutter2024quantum, zhu2022generative, zoufal2019quantum, situ2020quantum,zhang2025denoising}. Unfortunately, the diffusion processes dominated by the unitary evolution and the environmental perturbation remain unknown.

The explorations in quantum diffusion models demonstrate the diverse approaches for extending classical diffusion processes to the quantum domain. The first approach maintains classical diffusion while employing quantum ansatzs for denoising, leveraging quantum advantages in the learning process. The second approach utilizes quantum channels for the forward process, allowing for more complex noise dynamics while relying on classical neural networks for denoising. The third approach implements both phases in the quantum domain, fully exploiting quantum mechanical properties throughout the entire process.

To prepare the quantum state generation, early works focused on combining classical diffusion with quantum denoising circuits. Cacioppo et al. \cite{cacioppo2023quantum} demonstrated this hybrid approach by replacing neural networks with parameterized quantum circuits, achieving practical implementation on real quantum devices. De Falco et al. \cite{de2024quantum} explored the potential of quantum latent diffusion models, extending classical latent space concepts to the quantum domain. More recent works have explored fully quantum implementations. Zhang et al. \cite{zhang2024generative} proposed QuDDPM utilizing quantum stochastic circuits for noise injection and parameterized quantum circuits for denoising, while Chen et al. \cite{chen2024quantum} developed a framework based on non-unitary evolution and partial trace operations. These approaches demonstrate the potential benefits of leveraging quantum mechanics throughout the entire process, particularly in handling quantum noise and state evolution. A brief overview of the classical framework from a physical perspective is provided in Appendix~\ref{Diffusion Models in Physical Perspective}. 

Despite these advances, current quantum diffusion models face two critical challenges. First, they lack a rigorous physical interpretation—most approaches directly transplant classical diffusion mechanisms to quantum systems without considering the essential role of quantum decoherence and open system dynamics. Second, the backward denoising process requires appropriate quantum channels effectively reverse the forward diffusion maintaining complete positivity and trace preservation, a non-trivial constraint in quantum mechanics. To further advance, understanding quantum diffusion from a physical perspective is crucial. Unlike classical noise which represents statistical fluctuations, quantum noise arises from system-environment interactions and exhibits uniquely quantum features such as non-commutativity and entanglement generation. This distinction suggests that quantum diffusion models should be grounded in open quantum system theory rather than classical analogs.

To address these challenges, we develop a channel-constrained Markovian quantum diffusion (CCMQD) model that treats diffusion as the natural evolution of open quantum systems. The highlights of our work are demonstrated as follows: (1) Proposing a physical framework for quantum diffusion based on open quantum system dynamics, modeling the forward process as decoherence and the backward process as a learnable denoising channel; (2) Developing a channel-constrained methodology, based on Kraus representations, to learn physically-valid denoising processes for quantum state reconstruction; (3) Achieving high-fidelity $(0.999\pm1\times10^{-4})$ generation of complex quantum states, including $7$-qubit entangled states, across diverse noise models.

This work is organized as follows: in Section \ref{Quantum Generative Diffusion Model Based on open quantum systems}, we introduce our Channel-Constrained Markovian Quantum Diffusion (CCMQD) model, deriving its architecture from the principles of open quantum systems and detailing its constrained learning framework. In \ref{Experiments}, we present the numerical results, validating the model's performance, scalability, and robustness through extensive experiments and benchmarking. Finally, Section \ref{Discussion} concludes with a discussion of the physical insights and implications of our work.

\newpage

\section{Channel-Constrained Markovian Quantum Diffusion Model}
\label{Quantum Generative Diffusion Model Based on open quantum systems}

This section introduces our Channel-Constrained Markovian Quantum Diffusion (CCMQD) model. Although our approach is conceptually inspired by the classical diffusion framework (see Appendix~\ref{Diffusion Models in Physical Perspective}), its construction is grounded in rigorous physical foundations. To begin with, we establish this foundation based on the principles of open quantum systems. Subsequently, we employ the formal framework of quantum channel-defined models. Finally, we outline the constraint optimization and training framework required to implement this model.

\subsection{Physical Foundation}
\label{QME}

An open quantum system inevitably interacts its surrounding environment. This interaction is the physical origin of quantum noise and decoherence, leading to a loss of quantum information. The quantum master equation \cite{li2005quantum}, which serves as a standard formulation for characterizing the dynamics of open quantum systems, takes on the distinct form of the Lindblad equation \cite{pearle2012simple}, meticulously accounting for the interactions occurring with the environment. For an open quantum system, the total Hamiltonian can be decomposed as
\begin{equation}
	H = H_s + H_E + H_{int},
\end{equation}
where $H_s$ and $H_E$ represent the Hamiltonians of the quantum system and the environment, respectively. The interaction Hamiltonian $H_{int}$ describes the coupling between the quantum system and its environment. For Markovian quantum dynamics, the evolution must satisfy complete positivity for the map between the states of the quantum system at different times. Consequently, we obtain the general master equation in Lindblad form
\begin{equation}
	\label{Lindblad}
	\frac{d\rho}{dt}=-i[H_{s},\rho(t)] + \sum_{n=1}^{N} \left( \Gamma_{n}\rho(t)\Gamma_{n}^{\dagger} - \frac{1}{2}\Gamma_{n}^{\dagger}\Gamma_{n}\rho(t) - \frac{1}{2}\rho(t)\Gamma_{n}^{\dagger}\Gamma_{n} \right),
\end{equation}
On the righthand side, the first term describes the quantum coherence effects due to the internal Hamiltonian of the quantum system. The second term represents the effect of the environment on the system dynamics, with each $\Gamma_n$ corresponding to a different decoherence process. This formulation provides a complete description of the quantum state evolution under environmental interactions, naturally incorporating both unitary dynamics and dissipative processes. The Lindblad form ensures that the evolution preserves the essential properties of the density matrix, including hermiticity, trace preservation, and positive semi-definiteness, making it particularly suitable for describing quantum noise processes in diffusion models.

\begin{figure}[t]
	\centering
	\includegraphics[width=0.95\columnwidth, trim=3cm 2.4cm 1.7cm 2.5cm, clip]{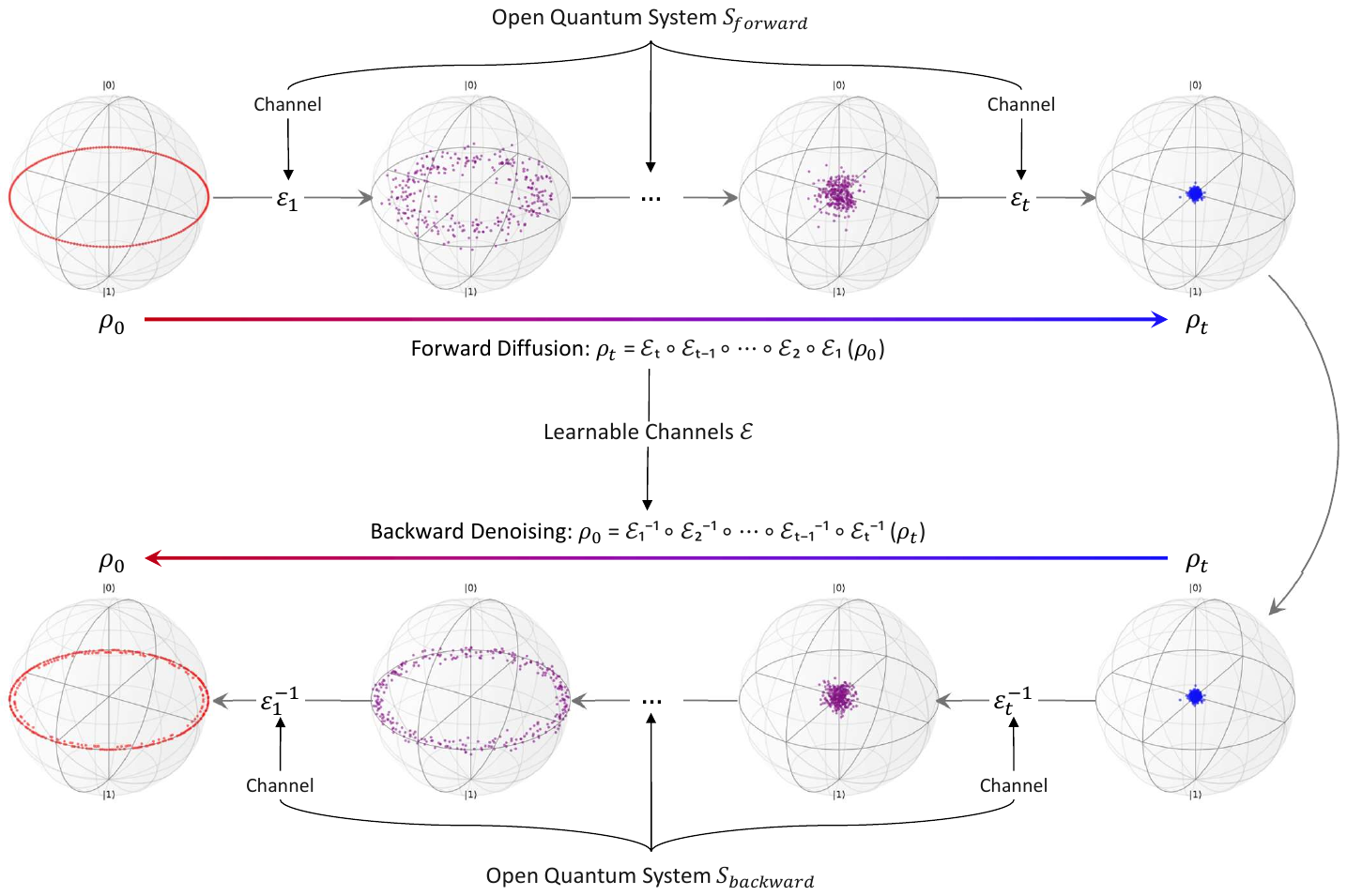}
	\caption{\textbf{Channel-based framework for quantum diffusion.} 
		Forward diffusion (top) progressively achieve diffusion through sequential quantum channels $\mathcal{E}_i$
		, while backward denoising (bottom) reconstructs the initial state through learned inverse channels $\mathcal{E}^{-1}_i$. Each channel is characterized by its Kraus operator decomposition.}
        \label{Fig:Channel_in_open_system}
\end{figure}

To translate this continuous-time physical process into a discrete step-by-step computational model, we approximate the evolution governed by the Lindblad equation (Eq.~\ref{Lindblad}) over a small time interval $\Delta t$. As derived in Eq.~\ref{C6} of Appendix~\ref{Channel-Based Unified Framework}, this discretization of the continuous Markovian evolution naturally leads to a discrete-time quantum channel $\mathcal{E}$, which is a completely positive trace-preserving (CPTP) map acting on the state. This single step of decoherence is thus described as $\rho_{t}=\mathcal{E}_{t}(\rho_{t-1})$. The entire forward diffusion is then a sequence of such channel operations (see Fig. \ref{Fig:Channel_in_open_system}).

This allows us to establish a unified, channel-based framework for quantum diffusion, as illustrated in Fig.~\ref{Fig:Channel_in_open_system}. The forward diffusion is modeled as a sequence of fixed noise channels $\{\mathcal{E}_t\}$ that represents the natural decoherence process. Conversely, the backward denoising process is achieved by learning a sequence of inverse quantum channels $\{\mathcal{E}_t^{-1}\}$ designed to systematically reverse the effects of decoherence and reconstruct the initial quantum state. This property of open quantum system dynamics builds up the fundamental framework of our designed CCMQD model.

\subsection{Model Architecture}
\label{Model Architecture}

The architecture of the CCMQD model implements the physical framework established in Sec.~\ref{QME}. This framework successfully reinterprets the forward diffusion and backward denoising processes as sequences of quantum channels, denoted as $\{\mathcal{E}_t\}$ and $\{\mathcal{E}_t^{-1}\}$, respectively. To implement this channel-based framework computationally, we must endow these abstract completely positive trace-preserving (CPTP) maps with a concrete mathematical form.

\begin{figure}[t]
	\centering
	\includegraphics[width=1\columnwidth, trim=5.5cm 4.5cm 2.7cm 5cm, clip]{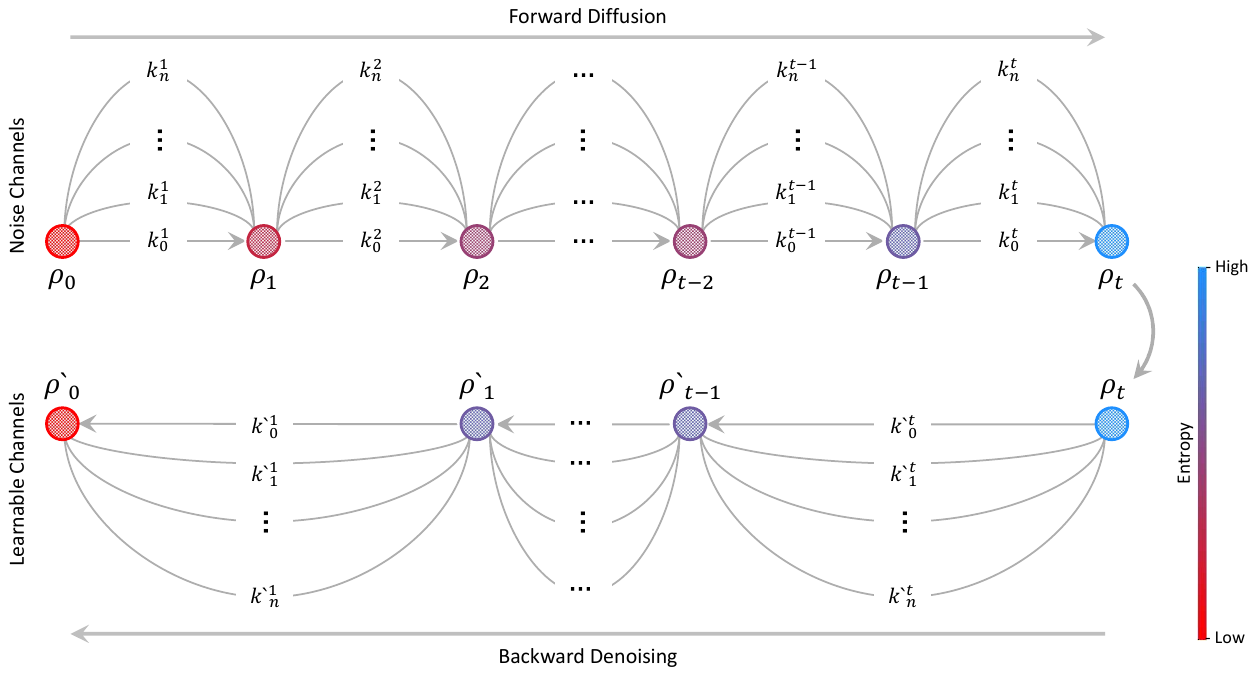}
	\caption{\textbf{Quantum diffusion framework with forward noising and backward denoising.} 
		Framework showing forward diffusion via noise channels (upper, left-to-right) and backward denoising via learnable channels (lower, right-to-left). Colors indicate entropy: red (low) to blue (high). Asymmetric time steps reflect different dynamics of noising and denoising processes.}
	\label{main Architecture}
\end{figure}

This mathematical form is not an ad hoc choice, but derives directly from the open quantum system dynamics that the channel represents. As formally detailed in Appendix~\ref{GQDP}, a quantum channel $\mathcal{E}_t$ is rigorously defined as the evolution of a system coupled to an environment. Equation~\ref{C3} in the appendix explicitly formalizes this, showing that the channel map is the result of a unitary evolution $U_t$ and a partial trace over the environmental degrees of freedom ($Tr_E$). Crucially, this operation is mathematically equivalent to the operator-sum representation, or Kraus decomposition, which is also shown in Eq.~\ref{C6}. This fundamental theorem allows us to define each channel by a set of Kraus operators $\{k_i^{(t)}\}$, which must satisfy the completeness relation $\sum_i (k_i^{(t)})^\dagger k_i^{(t)} = I$ to ensure the map is trace-preserving. 
Relying on the Kraus representation, we now define the overall architecture of the CCMQD model. As illustrated in Fig.~\ref{main Architecture}, the model consists of two symmetric but functionally distinct processes.

\begin{itemize}
	\item \textbf{The forward process} is implemented through a sequence of the fixed quantum noise channels $\{\mathcal{E}_t\}$. Each channel embodies specific quantum noise effects through its Kraus operator representation, progressively increasing the system's entropy and driving the state from its initial configuration $\rho_0$ towards a high-entropy, near-maximally mixed state $\rho_t$.
	\item \textbf{The backward process} is realized through a sequence of \textit{learnable} quantum noise channels $\{\mathcal{E}_t^{-1}\}$, where the Kraus operators are treated as trainable parameters. These parameters are constrained to preserve quantum mechanical properties, enabling systematic quantum state generation through controlled noise reduction and coherence restoration.
\end{itemize}

This two-step distinct physical dynamics requires a rigorous mathematical algorithm to describe the discrete-time and step-by-step evolution. The transitions between these states are governed by discrete-time CPTP maps, which are precisely our quantum channels $\mathcal{E}_t$ and $\mathcal{E}_t^{-1}$. This formalism provides a rigorous foundation for modeling the entire diffusion-denoising trajectory as a controlled Markovian evolution. We will now define the specific parameterization and mathematical formulation for each of these processes. The evolution of the sequence of density matrices $\{\rho_0, \rho_1, ..., \rho_t\}$ is determined by Kraus operators. Considering the Eq.~\ref{C4}, therefore, the Quantum Hidden Markov Process(QHMP) theory is partially applied.  The corresponding theoretical details of QHMP are given in Appendix~\ref{QHMP}.

\begin{figure}[t]
	\centering
	\includegraphics[width=1\columnwidth, trim=4.2cm 6.5cm 3.5cm 6cm, clip]{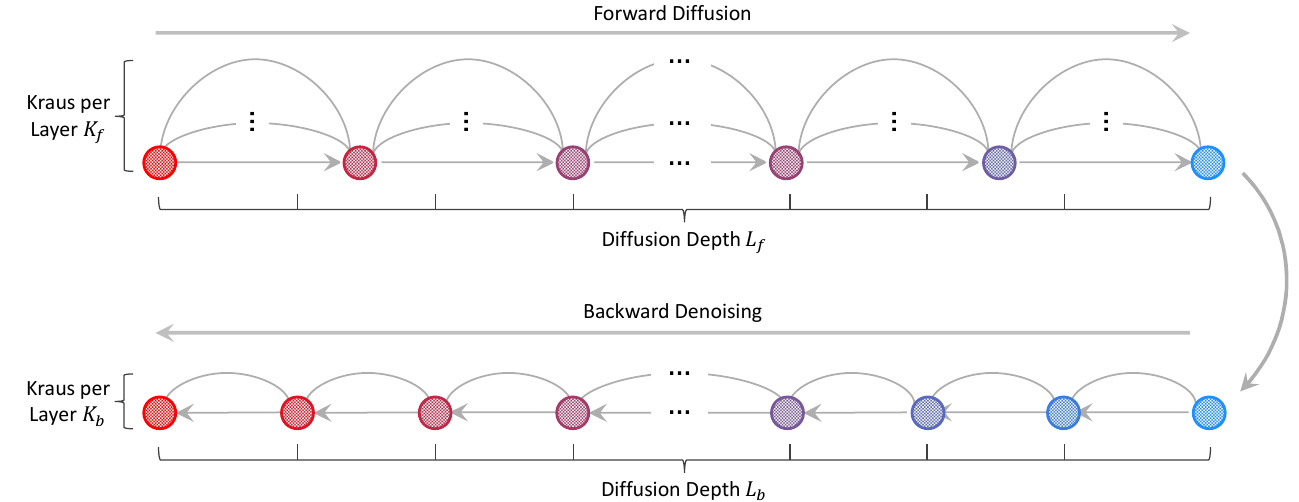}
	\caption{\textbf{Schematic representation of $(L_f, k_f)$ and $(L_b, k_b)$ noise configurations}}
	\label{Noise_Config}
\end{figure}

The implementation of CCMQD requires construction of quantum channels that respect physical constraints. For convenience, we parameterize the forward diffusion process as $(L_f, k_f)$, where $L_f$ denotes the diffusion depth and $k_f$ the number of Kraus operators per channel as shown in Fig.~\ref{Noise_Config}. Similarly, the backward denoising employs parameters $(L_b, k_b)$ for denoising depth and operator count, respectively.

For a quantum state $\rho_{t-1}$ at time $t-1$, the evolution is described as
\begin{equation}
	\label{channel_constraints}
	\rho_t = \mathcal{E}_t(\rho_{t-1}) = \sum_{i=1}^{k_f} k_i^{(t)} \rho_{t-1} (k_i^{(t)})^\dagger, \quad \sum_{i=1}^{k_f} (k_i^{(t)})^\dagger k_i^{(t)} = I
\end{equation}
The channel constraints are explicitly enforced through the completeness relation, ensuring trace preservation: $\text{Tr}[\mathcal{E}_t(\rho_{t-1})] = \text{Tr}[\rho_{t}]$. Eq.~\ref{channel_constraints} implements the quantum channel decomposition established in Appendix~\ref{GQDP}, where each $\mathcal{E}_t$ corresponds to an environmental interaction channel. The Kraus operators $\{k_i^{(t)}\}_{i=1}^{k_f}$ provide the explicit operator-sum representation of the forward diffusion channels, with $k_f$ denoting the number of Kraus operators per time step and the superscript $(t)$ indicating the time dependence of the environmental coupling strength. The detailed derivation of this channel-based description is provided in Appendix~\ref{GQDP}.

Building upon the composition of quantum channels in the forward process, the quantum state evolution can be further characterized through conditional probability relations. For an initial state $\rho_0$, while the sequential application of Kraus operators yields as
\begin{equation}
	\rho_t = \mathcal{E}_t \circ \mathcal{E}_{t-1} \circ \cdots \circ \mathcal{E}_1(\rho_0) = \sum_{i_1,\ldots,i_t} k^t_{i_t}\cdots k^1_{i_1}\rho_0(k^1_{i_1})^\dagger\cdots(k^t_{i_t})^\dagger.
\end{equation}
The concept of a simple transition probability does not directly apply to the deterministic evolution of density matrices. Instead, the relationship between the states at different times is quantified by the quantum fidelity, $F(\rho_a, \rho_b)$, which measures their similarity \cite{nielsen2001quantum}. For the forward process, the fidelity $F(\rho_t, \rho_{t-1})$ quantifies the information lost in a single step. For the entire diffusion process, the fidelity between the initial and final states is given by
\begin{equation}
	F(\rho_t, \rho_0) = F(\mathcal{E}_{t}\circ\cdot\cdot\cdot\circ\mathcal{E}_{1}(\rho_{0}), \rho_0).
\end{equation}
This fidelity naturally approaches $1/d$ as the state diffuses towards the maximally mixed state. As the diffusion depth increases to $L_{fmax}$, the quantum state naturally evolves toward a maximum entropy state, approximating the maximally mixed state $\rho_{L_{fmax}} \rightarrow \frac{1}{d}\mathbb{I}$, where $d$ denotes the dimension of the Hilbert space.
For the backward denoising process, we construct a sequence of learnable quantum channels with depth $L_b$ and $k_b$ Kraus operators per layer. The quantum state evolution during denoising is described as
\begin{equation}
	\rho_{t-1} = \mathcal{E}_t^{-1}(\rho_t) = \sum_{i=1}^{k_b} k_i^{\prime(t)} \rho_t (k_i^{\prime(t)})^\dagger, \quad \sum_{i=1}^{k_b} (k_i^{\prime(t)})^\dagger k_i^{\prime(t)} = I.
\end{equation}
Here, $\mathcal{E}_t^{-1}$ represents the learned inverse channels, where the trainable Kraus operators $\{k_i^{\prime(t)}\}_{i=1}^{k_b}$ approximate the inverse environmental interactions. The prime notation distinguishes the learned denoising parameters from the fixed forward diffusion operators, while $k_b$ allows for asymmetric channel structures where the denoising process may employ different numbers of Kraus operators than the forward process. Physically, these operators represent learned approximations to the inverse environmental interactions, with each $k_i^{\prime t}$ encoding a specific mechanism for coherence restoration. The backward denoising process, starting from the $\rho_{L_{f}}$, reconstructs the initial state through learnable quantum channels
\begin{equation}
	\rho_0 = \mathcal{E}_1^{-1} \circ \mathcal{E}_2^{-1} \circ \cdots \circ \mathcal{E}_{t}^{-1}(\rho_L) = \sum_{i_1,\ldots,i_{t}} k_{i_1}^{\prime(1)} \cdots k_{i_{L'}}^{\prime(L')} \rho_L (k_{i_{L'}}^{\prime(L')})^\dagger \cdots (k_{i_1}^{\prime(1)})^\dagger.
\end{equation}

where the sum over all Kraus operator combinations represents the ensemble of possible denoising pathways, each weighted by its quantum mechanical probability amplitude.

Based on the quantum state evolution relationships established above, a gradient-based learning process can be constructed to capture the transition between quantum states $\rho_t$ and $\rho_{t-1}$ at each diffusion depth. The complete denoising process emerges through the optimization of these transitions.

\subsection{Constrained Optimization and Training}
\label{subsec:training}

The primary challenge in training the backward channels is enforcing the physical constraints outlined in Section.~\ref{Model Architecture}. To learn the optimal set of physically-valid inverse channels, we developed the following constrained optimization and training framework.

\paragraph{Stiefel Manifold Optimization.}
The primary challenge in training the backward channels is enforcing the completeness relation, $\sum_{i}({k'}_{i}^{(t)})^{\dagger}{k'}_{i}^{(t)}=I$, at every step of the gradient-based optimization. We address this by representing the entire set of $N$ learnable Kraus operators as a single matrix $\kappa \in \mathbb{C}^{nN \times n}$, which is constrained to the Stiefel manifold $St(n, nN)$. This directly enforces the condition that defines the manifold $\kappa^{\dagger}\kappa=I$. The optimization aims to minimize a loss function $\mathcal{L}(\kappa)$. While the standard Euclidean gradient $G = \frac{\partial\mathcal{L}}{\partial\kappa}$ is computed, the parameter update cannot follow a simple step in this direction as it would violate the constraint. Instead, the gradient is projected onto the tangent space of the manifold, and the parameters are updated along the resulting geodesic. This procedure guarantees that the updated operators continue to satisfy $\kappa^{\dagger}\kappa=I$, ensuring a physically valid quantum channel throughout the training. The explicit update rules for this manifold optimization are detailed in Appendix~\ref{Stiefel Manifold Optimization}.

\paragraph{Channel-Constrained Training Strategies.}
We introduce and compare two distinct strategies for training the sequence of denoising channels. The first, Sequential Quantum Channel Optimization (SQCO), is a modular approach where each denoising channel $\mathcal{E}_t^{-1}$ is trained independently to reverse its corresponding forward step. The optimization objective at each step $t$ is to minimize a local loss function, defined by the fidelity between intermediate states:
\begin{equation}
	\mathcal{L}_{\text{SQCO}}^{(t)} = 1 - F(\rho_{t-1}, \hat{\rho}_{t-1}), \quad \hat{\rho}_{t-1} = \mathcal{E}_t^{-1}(\rho_t).
	\label{eq:sqco_loss}
\end{equation}
The second strategy, Holistic Quantum Trajectory Optimization (HQTO), is a global approach where all denoising channels $\{\mathcal{E}_t^{-1}\}_{t=1}^{L_b}$ are trained jointly. This method optimizes for the end-to-end reconstruction by minimizing a single global loss function:
\begin{equation}
	\mathcal{L}_{\text{HQTO}} = 1 - F(\rho_0, \hat{\rho}_0), \quad \hat{\rho}_0 = (\mathcal{E}_1^{-1} \circ \dots \circ \mathcal{E}_{L_b}^{-1})(\rho_{L_f}).
	\label{eq:hqto_loss}
\end{equation}
The HQTO strategy is designed to capture non-local quantum correlations across the entire process. A detailed comparison of these strategies is presented in Appendix~\ref{Channel-Constrained Training Strategies}.

\paragraph{Channel-Constrained Loss Functions.}
The training is guided by a loss function that measures the dissimilarity between the generated and target quantum states. The baseline for the HQTO strategy is an end-to-end fidelity loss, which exclusively considers the final reconstructed state $\hat{\rho}_0$ and the initial state $\rho_0$. While simple, this approach ignores the physical plausibility of the intermediate steps in the reconstruction trajectory. To provide denser supervision and ensure the entire path is physically meaningful, we introduce the path-constrained loss (PC-loss) function, formulated as:
\begin{equation}
	\mathcal{L}_{\text{path}} = \left(1 - F(\rho_{0},\hat{\rho}_{0})\right) + \lambda\sum_{t=1}^{L_b} \alpha_{t} \left(1-F(\rho_{t},\hat{\rho}_{t})\right).
	\label{eq:pc_loss}
\end{equation}
The first term in Eq.~\ref{eq:pc_loss} is the standard end-to-end fidelity loss. The second term serves as a path constraint, providing supervision at intermediate steps by penalizing the deviation between the denoised states $\hat{\rho}_t$ and their corresponding ground-truth states $\rho_t$ from the forward process. The hyperparameter $\lambda$ balances the importance of endpoint accuracy versus trajectory fidelity. This approach encourages the model to learn a physically consistent reversal of the decoherence process, rather than an arbitrary transformation. Further details are provided in Appendix~\ref{Channel-Constrained Loss Functions}.

\section{Results}
 \label{Experiments}

\subsection{Numerical Setup}
\label{subsec:setup}

The primary goal of our numerical experiments is to validate the performance, scalability, and robustness of the CCMQD model. The task is defined as the high-fidelity generation of pure quantum states, starting from a near-maximally mixed state. We focus on pure state generation as it presents the most stringent test case, given that pure states possess maximum quantum coherence and are highly susceptible to decoherence.

Validation is performed on n-qubit systems, with n ranging from 1 to 7. For the forward diffusion, we employ two distinct and physically motivated noise models to simulate the environmental decoherence. The first is a structured \textbf{depolarizing channels} with noise strength increasing linearly with the diffusion depth, modeling cumulative Markovian decoherence. The second is an unstructured \textbf{random channel}, where Kraus operators are sampled from the Haar measure, representing a more chaotic and challenging decoherence regime. The diffusion process under random conditions is more difficult to reverse, not only incorporating the noise variation over time with regular patterns characteristic of conventional methods, but also offering broader applicability.

The central metric for evaluating the model's performance is the quantum fidelity between the target pure state $\rho_0$ and the reconstructed state $\hat{\rho}_0$, defined as $F(\rho_0, \hat{\rho}_0) = \left(\text{Tr}\sqrt{\sqrt{\rho_0}\hat{\rho}_0\sqrt{\rho_0}}\right)^2$. All reported fidelities are the mean and standard deviation calculated over five independent experimental runs.

All numerical simulations presented in this work were implemented using the MindQuantum open-source quantum computing framework \cite{xu2024mindspore}, which provides a powerful environment for quantum computation simulation. The source code and data required to reproduce the numerical results reported in this paper are publicly available at \href{https://github.com/UESTC-YLH/CCMQDM/}{Code}.

\subsection{Comparison of Training Strategies}

First, we investigate which training strategy is most effective for reversing the decoherence process. The core question is whether a modular optimization of individual denoising steps is sufficient, or if a global optimization across the entire trajectory is necessary to accurately reconstruct multi-qubit quantum coherence. Answering this provides insight into the nature of information loss and recovery in the diffusion process.

To address this, we designed an experiment comparing three distinct strategies: Sequential Quantum Channel Optimization (SQCO), and HQTO enhanced with a path-constrained loss (PC-loss). The comparison was performed on systems of 1, 2, and 3 qubits. For the forward process, we used a fixed random noise configuration of $(L_f, K_f) = (10, 4)$, while the backward process was set to $(L_b, K_b) = (10, 10)$.

\begin{table}[t]
	\small
	\caption{\label{tab:performance_comparison}%
		Quantum coherence reconstruction using different optimization strategies.}
	\begin{ruledtabular}
		\renewcommand{\arraystretch}{1.3}
		\begin{tabular}{cccccc}
			\multirow{2}{*}{\textrm{Qubits}} & \multirow{2}{*}{\textrm{$(L_f, K_f)$}} & \multirow{2}{*}{\textrm{$(L_b, K_b)$}} & \multirow{2}{*}{\textrm{SQCO$^a$}} & \multicolumn{2}{c}{\textrm{HQTO+PC-loss}} \\
			\cline{5-6}
			& & & & \multicolumn{1}{c}{$\lambda=1$} & \multicolumn{1}{c}{$\lambda=0.02$} \\
			\colrule
			$1$ & $(10,4)$ & $(10,10)$ & $0.8465\pm2\times10^{-2}$ & $0.9997\pm5\times10^{-5}$ & $0.9999\pm3\times10^{-5}$ \\
			$2$ & $(10,4)$ & $(10,10)$ & $0.6123\pm3\times10^{-3}$ & $0.9987\pm6\times10^{-3}$& $0.9998\pm7\times10^{-5}$ \\
			$3$ & $(10,4)$ & $(10,10)$ & $0.6569\pm2\times10^{-2}$ & $0.9981\pm1\times10^{-3}$ & $0.9997\pm1\times10^{-4}$ \\
		\end{tabular}
	\end{ruledtabular}
	\vspace{1pt}
	\footnotesize\raggedright
	
	$^a$ Values represent converged fidelity (mean ± std) over five independent experiments.
\end{table}

The results, presented in Table~\ref{tab:performance_comparison}, reveal a stark performance difference. The modular SQCO approach achieves a moderate fidelity of $0.84$ for a single qubit, but its performance degrades catastrophically as system size increases, dropping to $0.65$ for a 3-qubit system. In sharp contrast, the global HQTO approach with path constraints maintains exceptionally high and stable fidelities, consistently exceeding $0.9997$ for all tested system sizes, even with minimal constraint weight ($\lambda = 0.02$).

This performance gap leads to a crucial insight: a modular, step-by-step optimization is insufficient because it fails to capture the non-local quantum correlations that persist throughout the diffusion trajectory. The superior performance of the holistic approach confirms that to successfully reconstruct quantum coherence, the entire diffusion-denoising process must be treated as a single, unified Markovian evolution. This validates the foundational principle of our model and establishes HQTO with path constraints as the optimal strategy for subsequent experiments.

\subsection{Performance and Scalability}

Having identified the optimal training strategy, we now evaluate the performance of the CCMQD model under more challenging conditions. The objective is twofold: to test the model's \textit{robustness} against physically distinct decoherence mechanisms and its \textit{scalability} for generating complex, multi-qubit entangled states. These tests are designed to demonstrate the practical viability and advantages of our physics-based approach.

\begin{figure}[t]
	\centering
	\includegraphics[width=1\columnwidth, trim=0cm 6.8cm 0cm 0cm, clip]{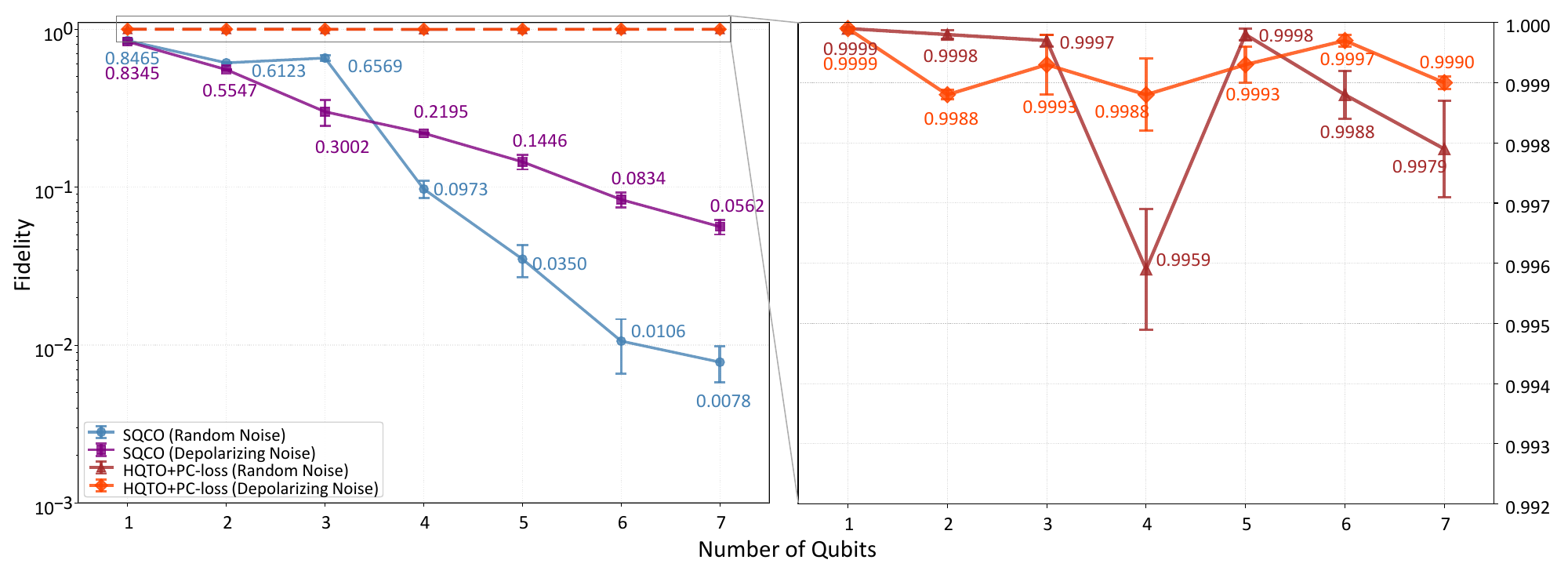}
	\caption{\textbf{Scalability of channel-constrained reconstruction across Quantum system scale.} Performance comparison between SQCO and HQTO+PC-loss for systems ranging from single qubits to 7-qubit entangled states. While SQCO shows rapid degradation with increasing system size, HQTO+PC-loss maintains fidelity $>0.99$  across systems under both noise types. Random noise conditions exhibit slightly reduced robustness compared to depolarizing noise, but still achieve reconstruction quality with fidelity $>0.997$.}
    \label{Scalability analysis}
\end{figure}

To test robustness, we benchmarked the model's state generation capability under two forward diffusion processes: one using structured depolarizing channels with linearly increasing noise strength, and another using unstructured channels with Kraus operators sampled from the Haar measure. To test scalability, we applied the model to the task of generating n-qubit entangled pure states, scaling the system size from $n=1$ to $7$. The results of these experiments are summarized in Table~\ref{tab:noise_comparison}, Table~\ref{tab:diffusion_depth_performance}, and Fig.~\ref{Scalability analysis}.

 \begin{table}[t]
	\small
	\caption{\label{tab:noise_comparison}%
		State generation under structured versus unstructured decoherence.}
	\begin{ruledtabular}
		\renewcommand{\arraystretch}{1.2}
		\begin{tabular}{ccccc}
			\textrm{Qubits} & \textrm{$(L_f, K_f)$} & \textrm{$(L_b, K_b)$} & \textrm{Random Noise} & \textrm{Depolarizing Noise} \\
			\colrule
			$1$ & $(10,4)$ & $(10,10)$ & $0.9999\pm3\times10^{-5}$ & $0.9999\pm9\times10^{-7}$ \\
			$2$ & $(10,4)$ & $(10,10)$ & $0.9998\pm7\times10^{-5}$ & $0.9988\pm8\times10^{-5}$ \\
			$3$ & $(10,4)$ & $(10,10)$ & $0.9997\pm1\times10^{-4}$ & $0.9993\pm5\times10^{-4}$ \\
			$4$ & $(10,4)$ & $(10,10)$ & $0.9959\pm1\times10^{-3}$ & $0.9988\pm7\times10^{-4}$ \\
		\end{tabular}
	\end{ruledtabular}
\end{table}

\begin{table}[t]
	\small
	\caption{\label{tab:diffusion_depth_performance}%
		Quantum state recovery from varying decoherence strengths.}
	\begin{ruledtabular}
		\renewcommand{\arraystretch}{1.1}
		\begin{tabular}{cccc}
			\textrm{Diffusion Depth} & \textrm{$(L_f, K_f)$} & \textrm{$(L_b, K_b)$} & \textrm{Fidelity} \\
			\colrule
			$6$ & $(6,4)$ & $(10,10)$ & $0.9991\pm6\times10^{-5}$ \\
			$5$ & $(5,4)$ & $(10,10)$ & $0.9990\pm5\times10^{-5}$ \\
			$4$ & $(4,4)$ & $(10,10)$ & $0.9974\pm8\times10^{-5}$ \\
			$3$ & $(3,4)$ & $(10,10)$ & $0.9997\pm4\times10^{-5}$ \\
			$2$ & $(2,4)$ & $(10,10)$ & $0.9998\pm3\times10^{-5}$ \\
		\end{tabular}
	\end{ruledtabular}
\end{table}

HQTO enhanced with PC-loss demonstrates remarkable performance and robustness across all scenarios. As shown in Tables~\ref{tab:noise_comparison} and \ref{tab:diffusion_depth_performance}, the reconstruction fidelity consistently exceeds 0.997 for both structured and unstructured noise types, even when the forward process is deep enough to evolve the initial state to a near-maximally mixed state. Furthermore, Fig.~\ref{Scalability analysis} illustrates the model's scalability; the fidelity remains above 0.997 as the system size increases to 7 qubits, showing no sign of degradation.

These results provide two key insights. First, the model's success against unstructured, Haar-random noise confirms that its effectiveness does not depend on exploiting a specific noise pattern, but rather on correctly enforcing the universal physical constraints of quantum channels. Second, the scalability demonstrates that the holistic training approach successfully preserves the complex quantum correlations present in exponentially large Hilbert spaces, a critical feature for generating entangled states that simpler, modular approaches fail to achieve.

\begin{table}[t]
	\small
	\caption{\label{tab:denoising_configurations}%
		Parameter efficiency in denoising channel configurations.}
	\begin{ruledtabular}
		\renewcommand{\arraystretch}{1.2}
		\begin{tabular}{cccc}
			\textrm{Qubits} &  \textrm{$(L_f, K_f)$} & \textrm{$(L_b, K_b)$} & \textrm{HQTO+PC-loss ($\lambda=0.02$)} \\
			\colrule
			$2$ & $(6,4)$ & $(6,10)$ & $0.9991\pm6\times10^{-5}$ \\
			$2$ & $(6,4)$ & $(6,9)$ & $0.9990\pm7\times10^{-5}$ \\
			$2$ & $(5,4)$ & $(5,10)$ & $0.9990\pm5\times10^{-5}$ \\
			$2$ & $(5,4)$ & $(5,9)$ & $0.9997\pm4\times10^{-5}$ \\
			$3$ & $(6,4)$ & $(6,10)$ & $0.9970\pm8\times10^{-5}$ \\
			$3$ & $(6,4)$ & $(6,20)$ & $0.9973\pm1\times10^{-3}$ \\
			$3$ & $(6,4)$ & $(12,16)$ & $0.9978\pm1\times10^{-3}$ \\
		\end{tabular}
	\end{ruledtabular}
\end{table}

To determine whether this strong performance relies on excessive parameterization, we examined the efficiency of the parameters used in the model. We tested various denoising configurations $(L_b, K_b)$ for multi-qubit systems, with the results summarized in Table~\ref{tab:denoising_configurations}. The data shows that increasing the complexity of the denoising channel — for instance, by doubling the number of Kraus operators for a 3-qubit system — yields only marginal gains in fidelity. This demonstrates that our framework is highly parameter-efficient, achieving near-optimal performance with a relatively compact denoising channel. The model's success is therefore rooted in its physically constrained structure rather than sheer parametric capacity.

\subsection{Benchmarking}

To contextualize the performance of the CCMQD model, we benchmark it against other prominent quantum generative models. We compare our model against two recent models: an Entangled Quantum Generative Adversarial Network (EQGAN) and the Randomised Quantum Generative Diffusion Model (RQGDM)~\cite{chen2024quantum}. The task for all models is the generation of n-qubit pure states from a near-maximally mixed initial state, with the system size n scaling from 1 to 7. This provides a direct comparison of scalability and performance on a challenging generative task.

\begin{table}[t]
	\small
	\caption{\label{tab:baseline_comparison}%
		Benchmarking against existing quantum generative models.}
	\begin{ruledtabular}
		\renewcommand{\arraystretch}{1.2}
		\begin{tabular}{cccc}
			\multirow{2}{*}{\textrm{Qubits}} & \multicolumn{3}{c}{\textrm{Quantum State Fidelity$^a$}} \\
			\cline{2-4}
			& \textrm{EQGAN\cite{chen2024quantum}} & \textrm{RQGDM \cite{chen2024quantum}} & \textrm{Ours work$^a$} \\
			\colrule
			$1$ & $0.972\pm3\times10^{-2}$ & --- & ${0.999\pm9\times10^{-7}}$ \\
			$2$ & $0.867\pm8\times10^{-2}$ & $0.999\pm6\times10^{-4}$ & ${0.998\pm8\times10^{-5}}$ \\
			$3$ & $0.741\pm1\times10^{-1}$ & $0.999\pm8\times10^{-4}$ & ${0.999\pm5\times10^{-4}}$ \\
			$4$ & $0.676\pm7\times10^{-2}$ & $0.996\pm8\times10^{-3}$ & ${0.998\pm6\times10^{-4}}$ \\
			$5$ & $0.602\pm6\times10^{-2}$ & $0.993\pm1\times10^{-3}$ & ${0.999\pm3\times10^{-4}}$ \\
			$6$ & $0.555\pm6\times10^{-2}$ & $0.990\pm1\times10^{-2}$ & ${0.999\pm1\times10^{-4}}$ \\
			$7$ & $0.463\pm1\times10^{-1}$ & $0.992\pm8\times10^{-2}$ & ${0.999\pm1\times10^{-4}}$ \\
		\end{tabular}
	\end{ruledtabular}
	\vspace{1pt}
	\footnotesize\raggedright
	
	$^a$ Bolding indicates the highest performance for the condition.\\
	$^b$ All methods set up forward processes long enough to ensure the generation of pure states from near-maximal mixing initial conditions.
\end{table}

In Table~\ref{tab:baseline_comparison} the results demonstrate that our CCMQD model consistently outperforms the other approaches, particularly as the system size increases. While the fidelity of the EQGAN model degrades significantly to $0.463$ for 7 qubits, and the RQGDM model achieves a high fidelity of $0.992$ but with a large variance, our model achieves a superior and highly stable fidelity of $0.999 \pm 1\times10^{-4}$.

This benchmark highlights the tangible benefits of our channel-constrained framework. The performance and stability, especially in the high-dimensional 7-qubit regime, suggest that modeling the generative process as a physically constrained Markovian evolution is a more robust and scalable strategy than existing adversarial or alternative diffusion-based approaches.

\subsection{Insights and Results}

The experimental results validate our central thesis: the success of the CCMQD model stems directly from its foundation in open quantum system dynamics. By treating diffusion as a controllable physical process rather than an abstract algorithm, we achieve a more robust and scalable approach to quantum state generation.

A key insight comes from the superior performance of the holistic training strategy (HQTO). The failure of the modular SQCO approach for multi-qubit systems demonstrates that quantum coherence reconstruction is a non-local problem. Successful denoising requires optimizing the entire quantum trajectory as a unified Markovian evolution, respecting the complex correlations that persist throughout the process, which is a direct consequence of the underlying quantum master equation.

This physical foundation marks a fundamental departure from previous models. Unlike classical diffusion, which requires full evolution to a simple prior distribution (e.g., Gaussian noise), our model can reverse decoherence from any intermediate state, as we have access to the complete density matrix at each step. This offers greater flexibility and potential applications in areas like quantum error mitigation. Furthermore, unlike prior quantum models that often adapt classical mechanisms without guaranteeing physical validity, our channel-constrained framework ensures every learned operation is a physically valid quantum channel (CPTP). This inherent physical consistency is the primary reason for the model's high fidelity and robust scalability, especially in high-dimensional Hilbert spaces where unconstrained methods falter. Moreover, the robust performance under random channels demonstrates that our method exhibits superior adaptability while avoiding potential dependence on regular noise patterns.

\section{Discussion and Conclusion}
\label{Discussion}

In this work, we have developed a channel-constrained Markovian quantum diffusion (CCMQD) model that provides a rigorous physical foundation for quantum generative modeling. By adapting the generative process to the open quantum systems, our model can describe forward diffusion as environmental decoherence and backward denoising as a learned, physically-valid reversal of this process. Our comprehensive validation on systems up to 7-qubit entangled states demonstrates the model's effectiveness, achieving state generation fidelities consistently exceeding 0.998 and showcasing robustness and scalability.

The success of this approach establishes a physical foundation for quantum diffusion that addresses two fundamental challenges: the lack of a rigorous physical interpretation and the difficulty of ensuring that learned operations are valid quantum channels. Our framework reveals that quantum diffusion is a natural consequence of open system dynamics governed by quantum master equations, shifting the perspective from a purely algorithmic procedure to a controllable physical process.

This physics-based perspective also uncovers fundamental distinctions from classical diffusion. Unlike classical models where intermediate states are only accessible through sampling, our framework provides the complete density matrix at any point in the forward process. This suggests a unique advantage: denoising can begin from an arbitrary intermediate state rather than requiring full diffusion to a simple prior, opening potential applications in quantum error mitigation. Furthermore, the von Neumann entropy is exactly computable at each step, enabling precise tracking of information loss and the potential for adaptive optimization of the diffusion process.

While our framework provides a complete theoretical description, it currently operates at the abstract quantum channel level. A crucial direction for future work is the compilation of these learned channels into efficient quantum gate sequences for implementation on near-term hardware. Photonic quantum systems, with their natural support for controllable environmental interactions, present a particularly promising platform for realizing this model. Ultimately, our results validate that treating quantum diffusion as a controlled Markovian evolution provides a robust and powerful path for generative modeling, suggesting a paradigm shift from simply circumventing environmental decoherence to actively harnessing it for quantum state synthesis.

\begin{acknowledgments}
This work was supported by the National Natural Science Foundation of China under Grant No. 62472072, and the CPS-Yangtze Delta Region IndustrialInnovation Center of Quantum and Information Technology-MindSpore Quantum Open Fund, and the Natural Science Foundation of Xinjiang Uygur Autonomous Region under Grant No. 2024D01A17 and No. 2023D01A63, and the Chengdu Key Research and Development Program under Grant No. 2025-YF08-00109-GX.
\end{acknowledgments}

\bibliography{apssamp}

\newpage
\appendix

\section{Classical diffusion models from a physical perspective}
\label{Diffusion Models in Physical Perspective}

This appendix reviews the physical and mathematical foundations of the classical diffusion models that motivated the quantum framework developed in the main text. We briefly outline the forward processes as noise-injection and backward processes as denoising. Understanding this classical foundation is helpful for appreciating the fundamental differences and novel aspects of the quantum approach presented in Section~\ref{Quantum Generative Diffusion Model Based on open quantum systems}.

From a physical perspective, diffusion models describe a complete process which a system evolves from an ordered state to a disordered state and then recovers through a reverse process \cite{luo2022understanding}. The forward diffusion process is viewed as a process which the system gradually loses information through interactions with the environment, and the entropy of the system progressively increasing towards a maximum entropy state. Conversely, the reverse diffusion process achieves entropy reduction through additional information input, seeking to reconstruct the initial system state.

Gaussian noise-based forward diffusion process \cite{croitoru2023diffusion} is currently the most widely used application. In the forward diffusion process, after sampling an initial data point $x_0$ from the actual data distribution, the system generates a series of intermediate states $x_1, x_2, ..., x_t$ (where $t$ represents the total number of steps in the diffusion process) by iteratively adding Gaussian noise. This process can be precisely described by the following mathematical form: 
\begin{equation}
	x_t = \sqrt{1 - \beta_t} \times x_{t-1} + \sqrt{\beta_t} \times z_t,
\end{equation}
where $z_t \sim \mathcal{N}(0, I)$. This process can be equivalently expressed as
\begin{equation}
	q(x_t|x_{t-1}) = \mathcal{N}(x_t; \sqrt{1-\beta_t}x_{t-1}, \beta_t\mathbf{I}),
\end{equation}
where the variance schedule $\{\beta_t\}_{t=1}^T$ controls the noise strength at each step. By defining $\alpha_t = 1 - \beta_t$ and $\bar{\alpha}_t = \prod_{s=1}^{t}\alpha_s$, we can directly sample $x_t$ from the initial state as
\begin{equation}
	x_t = \sqrt{\bar{\alpha}_t}x_0 + \sqrt{1-\bar{\alpha}_t}\epsilon, \quad \epsilon \sim \mathcal{N}(0, \mathbf{I}).
\end{equation}
This closed-form expression reveals that as $t \to T$, where $T$ is the final moment of diffusion, where  the signal-to-noise ratio $\bar{\alpha}_t/(1-\bar{\alpha}_t)$ approaches zero, transforming the data distribution into pure Gaussian noise.

The reverse diffusion process reconstructs the original data by learning to invert this noise injection. This process begins by sampling from a noise distribution, then gradually denoises the sample using a trained neural network to eventually obtain samples from the actual data distribution. The core equation of the reverse process can be represented as a Markov chain like
\begin{equation}
	p(x_{t-1}|x_t) = \mathcal{N}(x_{t-1}; \mu_\theta(x_t,t), \sigma_t^2I),
\end{equation}
where $\mu_\theta(x_t,t)$ is the mean predicted by a neural network parameterized by $\theta$, and $\sigma_t^2$ is the noise variance at step $t$. The neural network is trained to predict the mean of the previous step's distribution, taking as input the current noisy sample and the step index. The training process employs the negative log-likelihood of the data under the reverse process as the loss function: $\mathcal{L}(\theta) = -\sum_i \log p(x_{t-1}|x_t)$. The neural network parameters $\theta$ are optimized by minimizing this loss function, enabling the network to effectively denoise the data at each step of the reverse process. 

There is a correspondence between classical diffusion models and quantum systems. The stochastic noise in classical diffusion finds its quantum counterpart in quantum noise, while the entropy-increasing evolution towards maximum entropy states parallels the natural evolution of quantum states towards maximally mixed states under prolonged noise exposure. This intrinsic similarity has served as a fundamental basis for most current quantum adaptations of diffusion models \cite{zhang2024generative, chen2024quantum}.
This feature prompts us to explore quantum systems in a more in-depth form via an open quantum system.

Leveraging fundamental concepts from quantum master equations and quantum hidden Markov processes as shown in Appendix~\ref{QHMP}, we establish a theoretical foundation for our quantum diffusion framework, treating it as a natural manifestation of system-environment interactions.

\section{Quantum hidden Markov process}
\label{QHMP}

This appendix details the Quantum Hidden Markov Process (QHMP) \cite{srinivasan2018learning,adhikary2020expressiveness,markov2022implementation}, which provides the formal mathematical language for describing the discrete-time evolution of an open quantum system subject to measurements. The CCMQD model presented in the main textis a direct application of this underlying mathematical structure where the diffusion process is modeled as a sequence of discrete channel operations.

QHMP extends the classical Hidden Markov Model to the quantum domain, providing a sophisticated framework for describing discrete-time quantum evolution under measurement and environmental interactions. In a QHMP, a quantum system evolves through a sequence of hidden states, where the transitions between states are governed by quantum mechanical principles, and the observable outcomes are obtained through quantum measurements.

The mathematical framework of a QHMP can be formally defined using a set of parameters $\lambda = \{\rho(0), \Phi, k_\mu\}$. The quantum system begins in an initial state described by density matrix $\rho(0)$, and its evolution between discrete time steps is governed by a completely positive trace-preserving (CPTP) map $\Phi$, expressed as $\rho(t) = \Phi[\rho(t-\Delta t)]$. This map encapsulates both unitary evolution and dissipative processes, maintaining consistency with the Lindblad evolution in the continuous-time limit.

The measurement process, a crucial component of QHMP, is described by a set of operators ${M_\mu}$, where each $M_\mu$ corresponds to a possible measurement outcome. 
The probability of observing outcome $\mu$ at time $t$ (given the state at $t-\Delta t$) is given by the Born rule as

\begin{equation}
	p(\mu|\rho(t-\Delta t)) = Tr[M_{\mu}\rho(t-\Delta t)M_{\mu}^{\dagger}]
\end{equation}

The total evolution of the state, summing over all possible outcomes, is given by the CPTP map $\Phi$ as

\begin{equation}
	\rho(t) = \Phi[\rho(t-\Delta t)] = \sum_{j} M_{j}\rho(t-\Delta t)M_{j}^{\dagger}
\end{equation}

Following a measurement with outcome $\mu$, the quantum state updates according to

\begin{equation}
	\rho(t) = \frac{K_\mu\rho(t-\Delta t)K_\mu^\dagger}{Tr[\sum_jK_\mu\rho(t-\Delta t)K_\mu^\dagger]},
\end{equation}

reflecting the fundamental quantum mechanical principle of measurement-induced state collapse.

The QHMP framework naturally incorporates the quantum mechanical principles of superposition and measurement, making it distinctly different from its classical counterpart. The hidden states in QHMP are quantum states that cannot be directly observed without performing measurements, and the measurement process itself introduces fundamental quantum uncertainties and state disturbances. This quantum nature makes QHMP particularly suitable for analyzing quantum noise processes and state evolution in quantum diffusion models, as it provides a discrete-time framework that captures both the quantum state transitions and the measurement-induced state changes.
The aforementioned theoretical foundations provide us with critical tools and insights for constructing a universal framework. By integrating these elements through the open quantum system, we can fundamentally reinterpret diffusion models on a quantum level.

\section{Quantum diffusion as an open system process}
\label{GQDP}

This appendix provides a detailed theoretical elaboration of the concepts introduced in Sec.~\ref{Quantum Generative Diffusion Model Based on open quantum systems}. We formally establish the connection between the continuous dynamics of open quantum systems and the discrete channel-based framework of our CCMQD model. 
Here, we will take an in-depth look at the specific mechanisms by which forward and reverse processes are described as quantum channel sequences, and introduce the Kraus operator and representation as fundamental tools for their mathematical characterization.

As highlighted by Nguyen et al. \cite{nguyen2025diffusion} and Parigi et al. \cite{parigi2024quantum}, understanding the relationship between quantum noise and information diffusion could provide new insights into designing more effective quantum generative models. By reexamine the physical nature of quantum diffusion from the perspective of open quantum systems, quantum diffusion models can be describe as the evolution of quantum states under the influence of noise. From the perspective of open quantum systems, this process can be understood as the interaction between a quantum system and its environment: the environmental influence manifests as various forms of quantum noise, while the system evolution can be described by quantum channels. This open system-based description provides a generalized theoretical framework, as it directly corresponds to real diffusion processes in the physical world—where systems gradually reach equilibrium through environmental interactions. Within this framework, the quantum diffusion process can be represented by a sequence of quantum channels.

\subsection{Forward and Backward Processes} 

In the forward diffusion process, quantum systems gradually lose their quantum coherence through continuous interaction with the environment. Mathematically, given the initial density matrix $\rho_0$ of a quantum state, the system state $\rho_t$ at time $t$ can be represented as the result of a sequence of quantum channels as follow 
\begin{equation}
	\label{forward}
	\rho_t = \mathcal{E}_t(\rho_{t-1}) = \mathcal{E}_t \circ \mathcal{E}_{t-1} \circ \cdots \circ \mathcal{E}_2 \circ \mathcal{E}_1(\rho_0),
\end{equation}
where $\circ$ denotes the composition of quantum channels, i.e., $\mathcal{E}_t \circ \mathcal{E}_{t-1}(\rho) = \mathcal{E}_t(\mathcal{E}_{t-1}(\rho))$, representing sequential application of quantum operations. Each quantum channel $\mathcal{E}_t$ is a completely positive trace-preserving (CPTP) map that encapsulates the effects of noise on the quantum state. This interaction leads to a continuous increase in system entropy, with the quantum state gradually evolving towards a maximum entropy state. From a physical perspective, this process simulates the inevitable decoherence of quantum systems under environmental noise, ultimately resulting in the loss of quantum information.

The backward diffusion process reconstructs the original quantum state through a sequence of learned inverse quantum channels:
\begin{equation}
	\label{backward}
	\rho_0 = \mathcal{E}^{-1}_1 \circ \mathcal{E}^{-1}_2 \circ \cdots \circ \mathcal{E}^{-1}_{t-1} \circ \mathcal{E}^{-1}_t(\rho_t),
\end{equation}
where $\mathcal{E}^{-1}_t$ represents the inverse quantum channel at time step $t$. Unlike the forward process's entropy increase, the backward process requires external control mechanisms to systematically reduce system entropy and restore quantum coherence.

\subsection{Channel-Based Unified Framework} 
\label{Channel-Based Unified Framework}

In quantum diffusion models, noise manipulation constitutes the fundamental mechanism underlying both forward decoherence and backward reconstruction processes. From the channel perspective, quantum noise can be rigorously characterized as quantum channels — completely positive trace-preserving maps that encapsulate all possible environmental interactions. This channel-based description provides a unified mathematical language: environmental decoherence, measurement-induced disturbances, and controlled reconstruction operations can all be expressed as quantum channels with specific structural constraints. This unified channel-based perspective is illustrated in Figure~\ref{Fig:Channel_in_open_system} , which demonstrates how quantum diffusion naturally emerges from the interplay between system dynamics and environmental interactions in open quantum systems.

Fig.~\ref{Fig:Channel_in_open_system} illustrates the channel-based unified framework for quantum diffusion. In the forward process, the quantum state evolves as Eq.~\ref{forward}, where each quantum channel $\mathcal{E}_i$ emerges from the open system dynamics $\mathcal{S}_{\text{forward}}$. The Bloch spheres visualize the progressive evolution from pure state $\rho_0$ (red ellipse) toward the maximally mixed state $\rho_t$ (blue center point), representing entropy increase through environmental interactions. The backward denoising process reconstructs the initial state through Eq.~\ref{backward}, where the inverse channels $\mathcal{E}_i^{-1}$ are learnable transformations generated by $\mathcal{S}_{\text{backward}}$. The bidirectional structure emphasizes that both forward diffusion and backward denoising operate through quantum channels arising from open system interactions, with the key distinction that forward channels represent natural decoherence while backward channels must be learned to reverse the diffusion process.

From the perspective of open quantum systems, this reconstruction process can be understood as the controlled evolution of the system-environment composite. The quantum channel evolution takes the form as
\begin{equation}
	\label{C3}
	\mathcal{E}_t(\rho_{t-1}) = \text{Tr}_E[U_t\rho_{t-1} \otimes |e_0\rangle\langle e_0|U_t^\dagger] = \sum_k \mathcal{E}_k\rho_{t-1}\mathcal{E}_k^\dagger,
\end{equation}
where $U_t$ is the unitary evolution operator, $|e_0\rangle\langle e_0|$ represents the initial environmental state, $\text{Tr}_E$ denotes the partial trace over environmental degrees of freedom, and $\mathcal{E}_k=<e_k|U_t|e_0>$ is the $k$-th order channel characterizing the quantum noise with $\{|e_k\rangle\}$ forming an orthonormal basis for the environment. In each channel $\mathcal{E}_k$, the quantum state $\rho_{t-1}$ is randomly replaced with a new state $\rho_t^k = \frac{\mathcal{E}_k\rho_{t-1}\mathcal{E}_k^\dagger}{\text{Tr}(\mathcal{E}_k\rho_{t-1}\mathcal{E}_k^\dagger)}$ with probability $p_k = \text{Tr}(\mathcal{E}_k\rho_{t-1}\mathcal{E}_k^\dagger)$. In order to take into account the measurement-induced state changes, we include the projection operators $P_m$ into Eq.\ref{C3} and obtain
\begin{equation}
	\label{C4}
	\mathcal{E}_t(\rho_{t-1}) = \sum_m \text{Tr}_{E}[P_m U_t \rho_{t-1} \otimes \sigma U_t^{\dagger} P_m] = \sum_m \mathcal{E}_m\rho_{t-1}\mathcal{E}_m^{\dagger},
\end{equation}
where $\sigma = \sum_j q_j|j\rangle\langle j|$ represents the environmental state with probability amplitudes $q_j$ and orthonormal environmental basis $\{|j\rangle\}$. The $m$-th order outcome channel is written as $\mathcal{E}_m=\sqrt{q_j}<e_m|P_mU_t|e_j>$ with $\{|{e_m}>\}$ forming an orthonormal basis for the measurement apparatus degrees of freedom. The measured outcomes correspond to observable outputs while quantum state evolution corresponds to hidden state transitions. Namely, quantum measurements and environmental decoherence share the same mathematical structure, and both behave as quantum channels evoking state transitions and information loss. Due to the fact that actual quantum states can only be probed through measurements that perturb the resulting system state, this unified description forms a natural foundation for the analysis and optimization of quantum diffusion models using quantum hidden Markov process theory, as demonstrated in Appendix~\ref{QHMP}.

To implement quantum diffusion in discrete time steps, we need to connect the continuous Lindblad evolution described in Appendix~\ref{QME} to our channel-based scenario. For small time intervals $\Delta t$, the quantum evolution can be approximated by the discrete channel as\cite{nielsen2001quantum}
\begin{equation}
	\label{C6}
	\rho(t + \Delta t) = \sum_{j=0}^n k_j\rho(t)k_j^\dagger,
\end{equation}
where $k_j$ is the Kraus operator satisfying the completeness condition $\sum_{j=0}^nk^{\dagger}_j k_j=I$, $k_0 = I - i\tilde{H}_s\Delta t$ denotes the no-jump operator and the jump-operator $k_j = \Gamma_j\sqrt{\Delta t}$ for $j \in \{1...n\}$ is used to describe the quantum noise effects. The effective Hamiltonian $\tilde{H}_s = H_s - \frac{i}{2}\sum_{j=1}^n \Gamma_j^\dagger\Gamma_j$ is an non-Hermitian operator. By comparing the right-hand side of Eq.~\ref{C4} and Eq.~\ref{C6}, thus, it is easy to address that each Kraus operator can be used to represent one diffusion channel.

\section{Detailed optimization methods}

This appendix provides a detailed mathematical and algorithmic description of the constrained optimization and training framework summarized in Section.~\ref{subsec:training}. We present the explicit update rules for optimization on the Stiefel manifold, offer an detailed description of the channel-constrained training strategies, and provide the formal definitions of the fidelity-based and path-constrained loss functions used in our model.

Unlike unconstrained optimization in classical diffusion models, our approach must ensure that all learned transformations remain valid quantum channels — preserving complete positivity and trace at every optimization step. The challenge lies not only in learning any inverse transformation, but also in exploring channels capable of high-fidelity reconstruction of quantum states within the restricted space of the CPTP mapping. The optimization methods presented here are specifically designed to navigate this constrained space, ensuring that the learning process operates entirely within the realm of physically realizable quantum operations while achieving effective denoising performance.

\subsection{Stiefel Manifold Optimization}
\label{Stiefel Manifold Optimization}

The channel constraints fundamentally determine the optimization geometry. Since Kraus operators must satisfy $\sum_i k_i^\dagger k_i = I$ to preserve trace, the optimization must be restricted to the Stiefel manifold. The set of trainable Kraus operators $\{k_i^{\prime(t)}\}$ at each denoising step can be collectively represented as a matrix $\kappa \in \mathbb{C}^{nN \times n}$ on the Stiefel manifold, satisfying $\kappa^\dagger \kappa = I$, where $N = \sum_{t=1}^{L'} K'$ denotes the total number of Kraus operators across all denoising steps and $n = d$ corresponds to the Hilbert space dimension. This parameterization ensures that the learned channels $\{\mathcal{E}_t^{-1}\}$ maintain the complete positivity and trace-preservation properties. The Quantum-Constrained Gradient Descent follows the manifold optimization procedure, which is common in intermediate state calculations of quantum hidden Markov models \cite{li2024new, adhikary2020expressiveness}. Given a gradient $G$ with respect to parameters $\kappa$, the constrained update from an initial solution $\kappa_0$ as follows
\begin{equation}
	G = \frac{\partial \mathscr{L}}{\partial \kappa},
\end{equation}
\begin{equation}
	\kappa = \kappa_0 - \tau \mathbf{U}(I + \frac{\tau}{2}\mathbf{V}^\dagger\mathbf{U})^{-1}\mathbf{V}^\dagger\kappa_0,
\end{equation}
where $\mathbf{U} = [G,\kappa_0]$, $\mathbf{V} = [\kappa_0, -G]$, and $\tau$ represents the learning rate. This update ensures that the trajectory remains on the Stiefel manifold, preserving the completeness condition $\sum_i {k'}^{t\dagger}_i {k'}^t_i = I$ throughout the optimization process. 

The loss function for training the quantum diffusion model is designed to maximize the similarity between the initial quantum state and the denoised state. For a quantum state $\rho_0$ and its corresponding denoised state $\hat{\rho}_0$, the basic loss function can be formulated using quantum fidelity as follow
\begin{equation}
	\mathscr{L}_0 = -\text{Tr}(\sqrt{\sqrt{\rho_0}\hat{\rho}_0\sqrt{\rho_0}}).
\end{equation}
Fidelity-based loss function provides a natural quantum mechanical measure of state similarity, analogous to the mean squared error used in classical diffusion models. While this basic formulation ensures end-to-end state recovery, more sophisticated loss designs incorporating intermediate denoising steps can be constructed.

\begin{algorithm}[H]
	\caption{Channel-Constrained Markovian Quantum Generative Diffusion Model}
	\label{alg:quantum_diffusion}
	\setstretch{1.1}
	\footnotesize
	\begin{algorithmic}[1]
		\STATE \textbf{Input:} Initial quantum state $\rho_0$, diffusion depth $L$, denoising depth $L'$
		\STATE \textbf{Parameters:} Forward Kraus operators $\{k^t_i\}_{i,t=1}^{K,L}$, trainable backward Kraus operators $\{{k'}^t_i\}_{i,t=1}^{K',L'}$
		
		\STATE \textbf{Forward Diffusion Process:}
		\FOR{$t = 1$ to $L$}
		\STATE $\rho_t = \mathcal{E}_t(\rho_{t-1}) = \sum_{i} k^t_i \rho_{t-1} (k^t_i)^\dagger$
		\ENDFOR
		\STATE Final noisy state: $\rho_L = \mathcal{E}_L \circ \mathcal{E}_{L-1} \circ \cdots \circ \mathcal{E}_1(\rho_0)$
		
		\STATE \textbf{Backward Denoising Process:}
		\FOR{$t = L'$ down to $1$}
		\STATE $\hat{\rho}_{t-1} = \mathcal{E}_t^{-1}(\hat{\rho}_t) = \sum_{i} {k'}^t_i \hat{\rho}_t ({k'}^t_i)^\dagger$
		\ENDFOR
		\STATE Reconstructed state: $\hat{\rho}_0 = \mathcal{E}_1 \circ \mathcal{E}_2 \circ \cdots \circ \mathcal{E}_{L'}(\rho_L)$
		
		\STATE \textbf{Quantum-Constrained Optimization:}
		\STATE Initialize Kraus matrix $\kappa \in \mathbb{C}^{nN \times n}$ on Stiefel manifold: $\kappa^\dagger \kappa = I$
		\WHILE{not converged}
		\STATE Compute fidelity loss: $\mathcal{L}_0 = -\text{Tr}\left(\sqrt{\sqrt{\rho_0}\hat{\rho}_0\sqrt{\rho_0}}\right)$
		\STATE Compute gradient: $G = \frac{\partial \mathcal{L}_0}{\partial \kappa}$
		\STATE Set matrix blocks: $\mathbf{U} = [G, \kappa_0]$, $\mathbf{V} = [\kappa_0, -G]$
		\STATE Update on Stiefel manifold: 
		\STATE \quad $\kappa \leftarrow \kappa_0 - \tau \mathbf{U}\left(I + \frac{\tau}{2}\mathbf{V}^\dagger\mathbf{U}\right)^{-1}\mathbf{V}^\dagger\kappa_0$
		\STATE Check convergence: $\|\mathcal{L}_0^{(k)} - \mathcal{L}_0^{(k-1)}\| < \epsilon$
		\ENDWHILE
		
		\STATE \textbf{Output:} Optimized quantum state $\hat{\rho}_0$ with fidelity $F(\rho_0, \hat{\rho}_0)$
	\end{algorithmic}
\end{algorithm}

\subsection{Channel-Constrained Training Strategies}
\label{Channel-Constrained Training Strategies}

The quantum diffusion process can be approached from different physical perspectives, leading to distinct training methodologies that reflect fundamental principles in quantum dynamics. We propose two approaches for training quantum diffusion models: Sequential Quantum Channel Optimization (SQCO) and Holistic Quantum Trajectory Optimization (HQTO).

\begin{figure}[t]
	\centering
	\includegraphics[width=1\columnwidth, trim=4.2cm 7cm 3.8cm 7cm, clip]{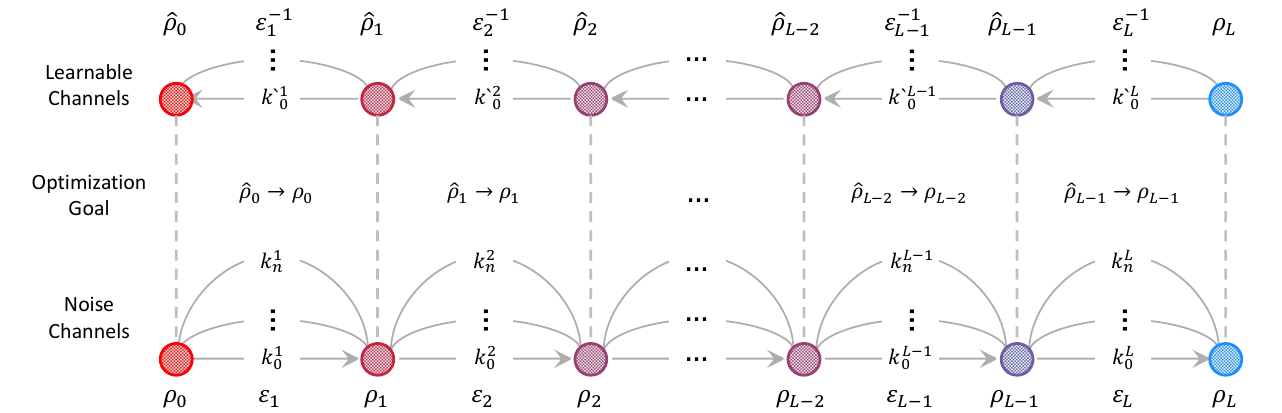}
	\caption{\textbf{SQCO optimization targets: step-wise quantum state reconstruction.} The optimization goal is to minimize the sum of differences between reconstructed states $\hat{\rho}_i$ and original states $\rho_i$ as $ \sum_{i=0}^{L-1} [1 - F(\rho_i, \hat{\rho}_i)]$, where $F$ denotes the fidelity. This follows the principle of modular optimization where individual denoising operations are specialized for their corresponding forward noise channels.}
    \label{SQCO optimization targets}
\end{figure}

Sequential quantum channel optimization (SQCO) uses the constraint structure of the channel by optimizing each $\mathcal{E}_i^{-1}$ independently while maintaining its individual CPTP properties. This approach ensures that each learned inverse channel satisfies the completeness relation locally, simplifying constraint enforcement. SQCO decomposes the complete quantum trajectory into a sequence of independent quantum channels, each denoising channel $\mathcal{E}^{-1}_t$ being trained separately to counteract its corresponding diffusion channel $\mathcal{E}_t$. The physical intuition behind SQCO aligns with the way quantum systems interact with their environment through local operations, is similar to that of quantum error correction protocols handling individual error channels.
After independent optimization of each layer-specific denoising channel, the complete denoising process is constructed by concatenating these channels in reverse order as $\hat{\rho}_0 = \mathcal{E}_1 \circ \mathcal{E}_2 \circ\cdots \circ \mathcal{E}_L(\rho_L)$.

\begin{figure}[t]
	\centering
	\includegraphics[width=1\columnwidth, trim=4.2cm 7cm 3.8cm 7cm, clip]{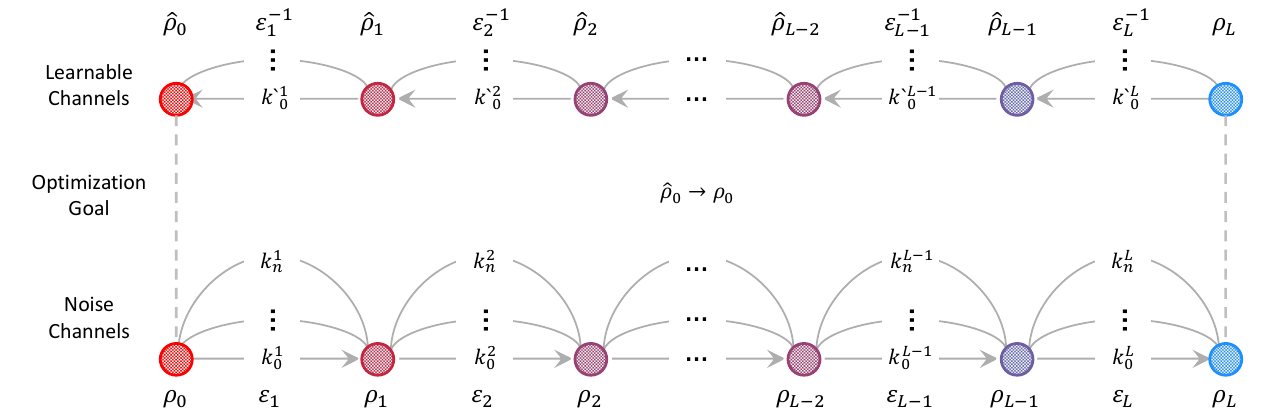}
	\caption{\textbf{HQTO optimization target: end-to-end quantum state reconstruction.} The optimization goal is to minimize the difference between the final reconstructed state $\hat{\rho}_0$ and original state $\rho_0$ as $[1 - F(\rho_0, \hat{\rho}_0)]$, where $F$ denotes the fidelity. This follows the principle of holistic optimization where all denoising channels are jointly trained to capture non-local quantum correlations across the entire diffusion trajectory.}
    \label{HQTO optimization target}
\end{figure}

Holistic Quantum Trajectory Optimization (HQTO) enforces channel constraints globally throughout the diffusion trajectory. While each $\mathcal{E}_i^{-1}$ must individually satisfy CPTP requirements, HQTO additionally ensures that their composition $\mathcal{E}_1^{-1} \circ \mathcal{E}_2^{-1} \circ \cdots \circ \mathcal{E}_L^{-1}$ remains a valid quantum channel, capturing non-local correlations that arise from the Markovian evolution. HQTO treats the entire quantum evolution as a unified process, acknowledging that quantum correlations can span across multiple time steps in the diffusion trajectory. All denoising channels $\{\mathcal{E}^{-1}_t\}_{t=1}^L$ are trained simultaneously to optimize the end-to-end fidelity between the original quantum state $\rho_0$ and the completely reconstructed state $\hat{\rho}_0$. The HQTO approach acknowledges that environmental interactions in open quantum systems can induce correlations across multiple time scales, making a global optimization strategy.

In summary, SQCO provides a modular approach where each denoising channel can be specialized for a particular noise type, potentially offering better interpretability of the learned quantum operations. HQTO can capture subtle interdependencies between diffusion steps that might be missed when optimizing channels independently.

\subsection{Channel-Constrained Loss Functions}
\label{Channel-Constrained Loss Functions}

The basic fidelity-based loss function introduced in Section~\ref{Stiefel Manifold Optimization} provides a quantum mechanical measure for evaluating the similarity between the original and reconstructed quantum states. However, this approach solely focuses on the endpoint states without considering the intermediate quantum states throughout the diffusion process. From the perspective of open quantum systems, the entire quantum trajectory contains valuable information about the evolution of the system in the presence of environmental interactions.

The Path-Constrained Loss (PCL) function directly incorporates channel constraints into the optimization objective \cite{chen2024quantum}. By enforcing fidelity at intermediate steps, we ensure that each channel $\mathcal{E}_i^{-1}$ not only preserves quantum mechanical validity but also correctly inverts the corresponding forward channel $\mathcal{E}_i$ within the constraints of CPTP maps. PCL aligns with the physical interpretation of denoising as the reverse process of diffusion in open quantum systems. PCL is formulated as
\begin{equation}
	\mathcal{L}_{\text{path}} = -\text{Tr}\left(\sqrt{\sqrt{\rho_0}\hat{\rho}_0\sqrt{\rho_0}}\right) + \lambda \sum_{t=1}^{L} \alpha_t \left(1 - \text{Tr}\left(\sqrt{\sqrt{\rho_t}\hat{\rho}_t\sqrt{\rho_t}}\right)\right),
\end{equation}
where $\rho_t$ represents the quantum state at step $t$ of the forward diffusion process, $\hat{\rho}_t$ denotes the corresponding state at the same relative position in the backward denoising process, $\lambda$ is a hyperparameter that balances the endpoint fidelity with path constraints, and $\alpha_t$ is the time-dependent weight for different diffusion stages.

The additional path constraints provide supervision signals at multiple points along the quantum trajectory decomposing the complex endpoint-matching problem into a sequence of incremental denoising steps. By constraining the entire trajectory, furthermore, the model is encouraged to learn physically plausible quantum noise reduction mechanisms rather than arbitrary transformations that violate the principles of quantum evolution due to the environmental interaction.

\section{Visualization}
\subsection{State Evolution Trajectory Visualization}
\label{Visualization1}

To supplement the fidelity-based analysis in the main text in Section~\ref{Experiments}, this appendix provides a detailed visualization of the quantum state's physical trajectory. We illustrate the single-qubit state's path on the Bloch sphere, offering a direct geometric intuition  for the decoherence and state reconstruction dynamics.

\begin{figure}[t]
	\centering
	\includegraphics[width=1\columnwidth, trim=4cm 0cm 4cm 0cm, clip]{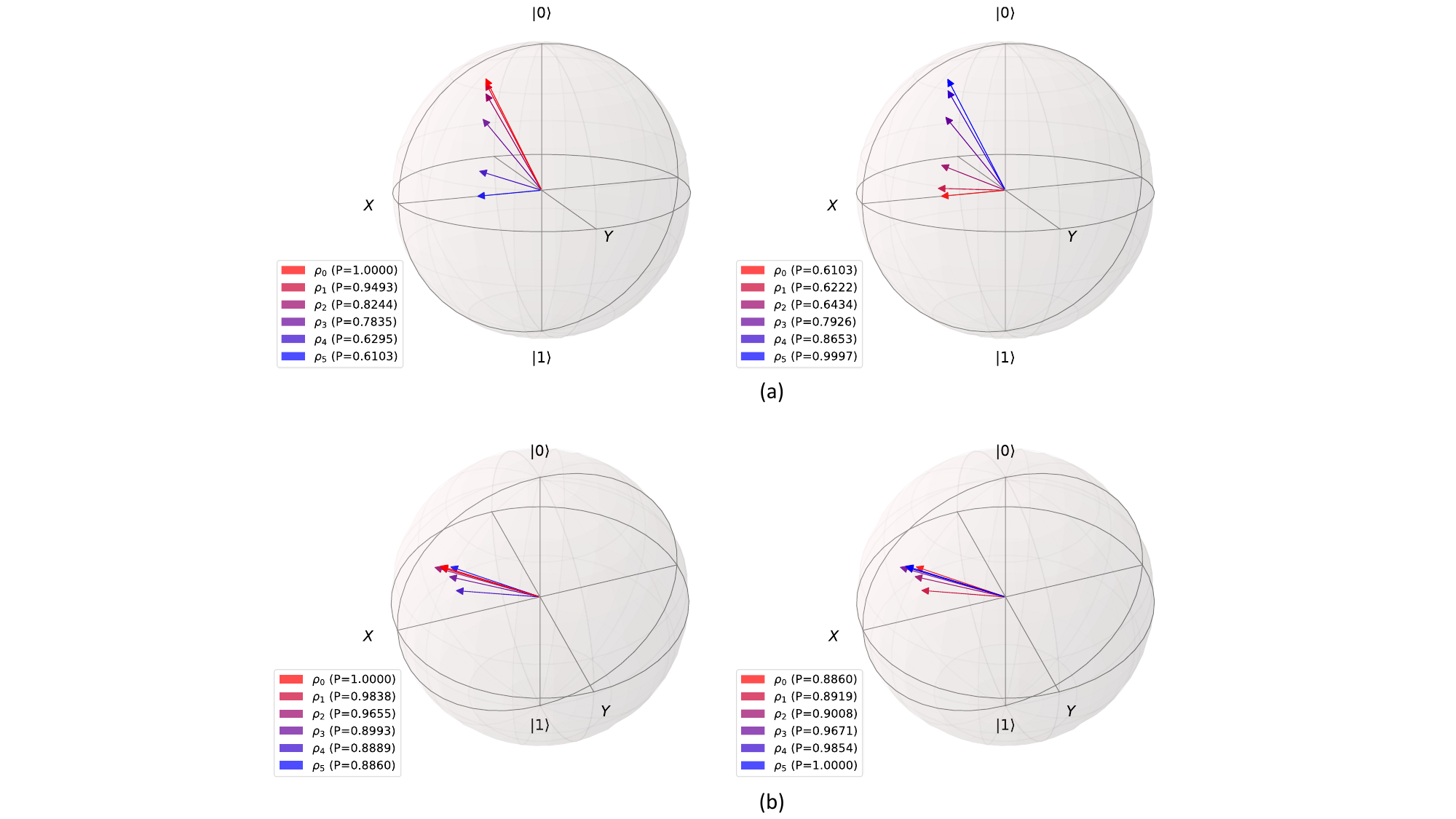}
	\caption{{\textbf{Bloch sphere visualization for a short diffusion process ($L=6$) under the HQTO+PC-loss.} The left plot in each pair shows the forward process, while the right plot shows the learned backward process. The trajectory direction is indicated by color: red marks the starting point (e.g., $\rho_0$) and blue marks the ending point (e.g., $\rho_5$). Each point's purity $P = Tr(\rho^2)$ is noted. (a) Evolution under depolarizing noise. (b) Evolution under random noise.}}
	\label{fig:bloch_evo_6}
\end{figure}

\begin{figure}[t]
	\centering
	\includegraphics[width=1\columnwidth, trim=4cm 0cm 4cm 0cm, clip]{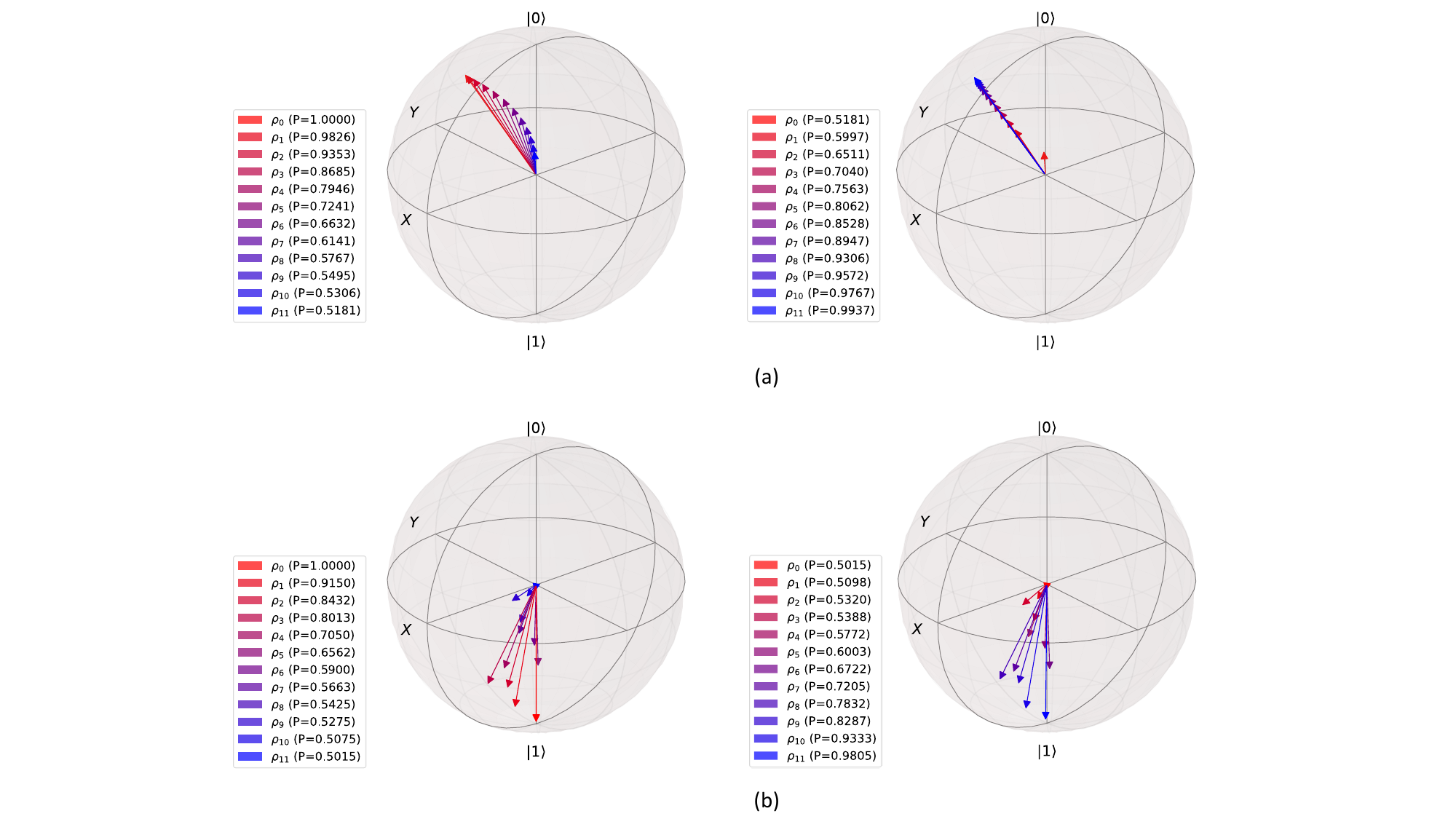}
	\caption{{\textbf{Bloch sphere visualization for a long diffusion process ($L=12$) under the HQTO+PC-loss.} The description follows Fig.~\ref{fig:bloch_evo_6}. (a) Evolution under depolarizing noise. (b) Evolution under random noise.}}
	\label{fig:bloch_evo_12}
\end{figure}

Figs.~\ref{fig:bloch_evo_6} and \ref{fig:bloch_evo_12} geometrically depict the CCMQD model's performance on a randomly selected initial pure state ($P=1.0000$). Each point on the path represents the quantum state at a discrete step, with its purity $P = Tr(\rho^2)$ explicitly noted. Several key insights, consistent with our model's physical foundation, can be extracted from this visualization. In most scenarios, particularly in the random noise cases (Fig.~\ref{fig:bloch_evo_6}(b) and Fig.~\ref{fig:bloch_evo_12}(b)), the learned backward process successfully reverses the quantum state's evolution along a path that closely mirrors the forward trajectory. This observation aligns with the objective of our HQTO+PC-loss strategy, which is designed to constrain the entire reconstruction path rather than just the endpoints.

However, a notable example is the right panel of Fig.~\ref{fig:bloch_evo_12}(b), the learned denoising trajectory clearly deviates from a simple time-reversal of the forward path, appearing to curve around the interior. This suggests that for certain noise models, particularly structured ones that drive the system to high entropy, the optimal learned channel may discover a valid reconstruction pathway that is not a simple mirror of the decoherence trajectory. The model may converge to a different, yet equally valid, solution for coherence restoration.

A crucial observation, validated across all scenarios, is the model's ability to initiate the denoising process from an intermediate, incompletely mixed state, which is particularly evident in the shorter diffusion processes depicted in Fig.~\ref{fig:bloch_evo_6}(a) and (b). Unlike many classical diffusion models, our framework does not require the forward process to evolve fully to a simple, known prior distribution to be effective. We have observed that numerous quantum diffusion models exhibit similar properties \cite{huang2025continuous,chen2024quantum}, regardless of whether time steps are incorporated into the learning process—this can be tentatively regarded as a universal characteristic of quantum diffusion models.

\subsection{Training Dynamics and Path Fidelity Convergence}
\label{Visualization2}

\begin{figure}[t]
	\centering
	\label{fig:fidelity_training}
	\includegraphics[width=1\columnwidth, trim=2.5cm 0.4cm 2cm 0cm, clip]{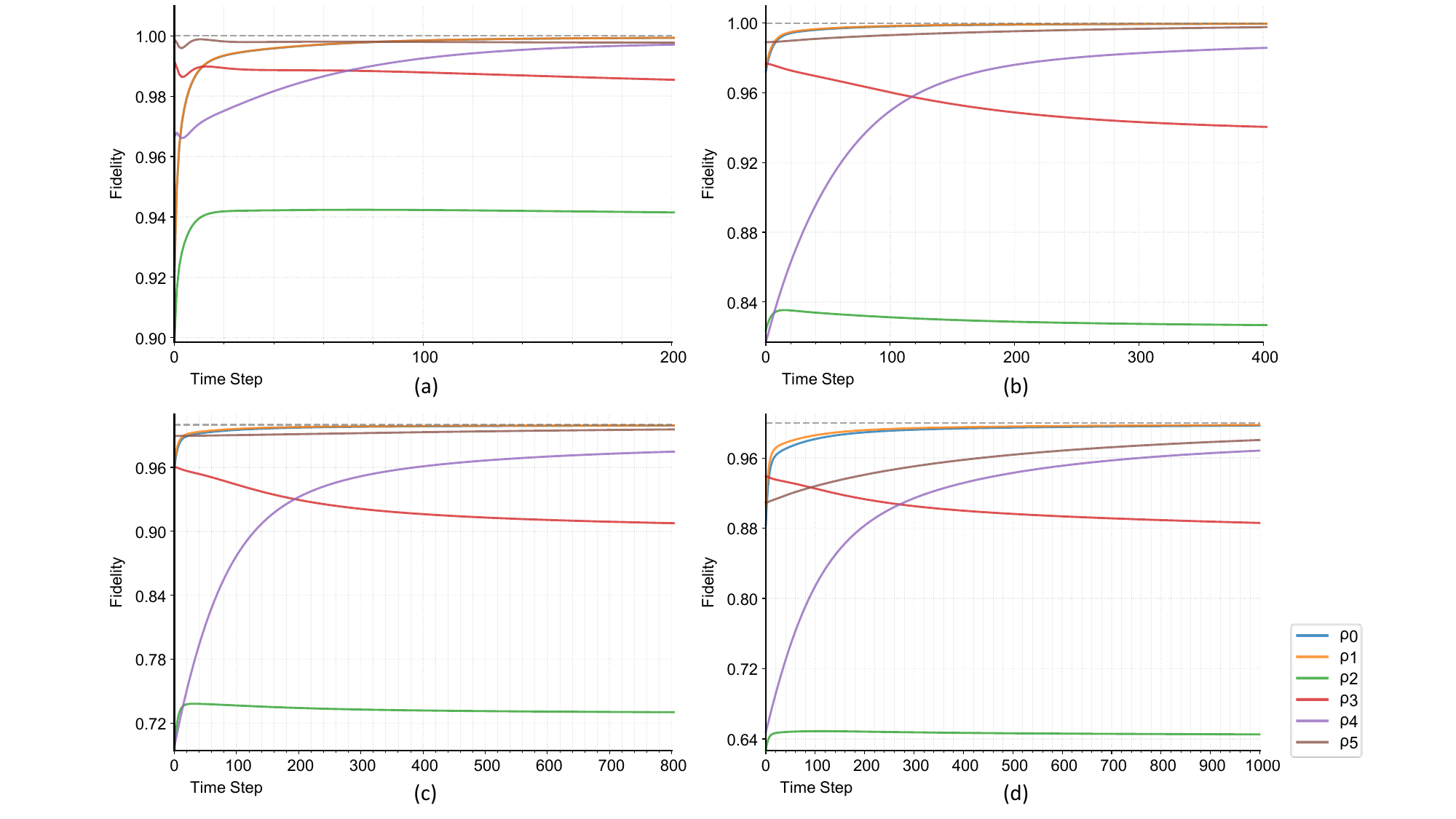}
	\caption{{\textbf{Convergence of intermediate state fidelities during HQTO+PC-loss.} The plots show the fidelity $F(\rho_t, \hat{\rho}_t)$ for all six states in the denoising path (labeled $\rho_0$ to $\rho_5$) as a function of training time steps. The system size scales as: (a) 1 qubit, (b) 2 qubits, (c) 3 qubits, and (d) 4 qubits. The final achieved fidelities for the target state ($\rho_0$) are $0.99980$ (a), $0.99985$ (b), $0.99941$ (c), and $0.99694$ (d)}}
\end{figure}

\begin{figure}[t]
	\centering
	\label{fig:loss_training}
	\includegraphics[width=1\columnwidth, trim=4cm 0.3cm 4cm 0cm, clip]{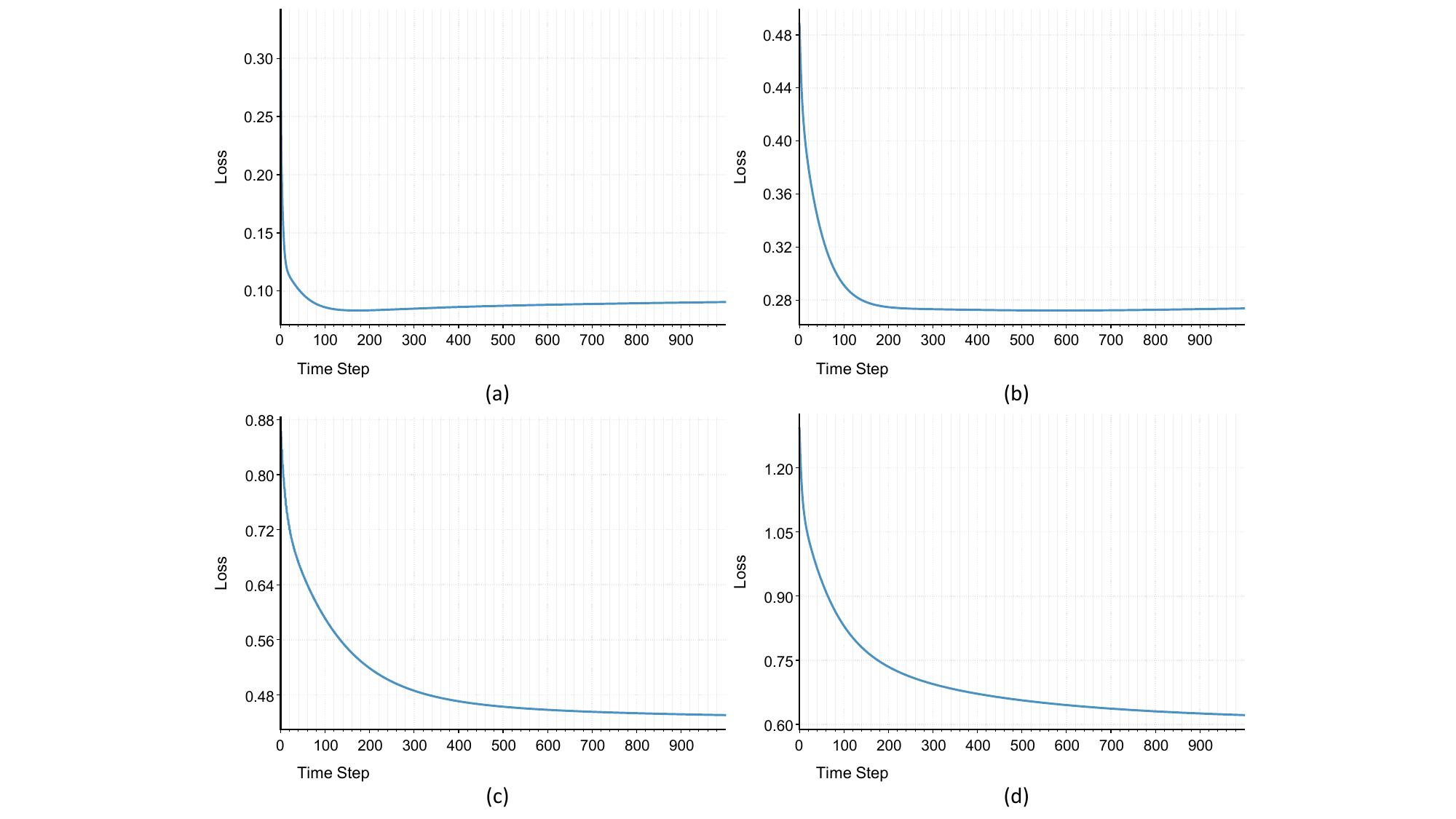}
	\caption{{\textbf{Convergence of the loss corresponding to the training runs in Fig.~\ref{fig:fidelity_training}.} The subplots show the loss function versus training time steps for: (a) 1 qubit, (b) 2 qubits, (c) 3 qubits, and (d) 4 qubits. All four scenarios demonstrate rapid initial loss reduction and stable convergence.}}
\end{figure}

Following the analysis of the final state trajectories, we now examine the training dynamics responsible for these results. This section details the convergence behavior of our model under the HQTO+PC-loss. We analyze the simultaneous evolution of the overall loss function (defined in Eq.~\ref{eq:pc_loss}) and the fidelities of the intermediate states, $F(\rho_t, \hat{\rho}_t)$, throughout the training process. The analysis is presented for systems scaling from $n=1$ to $n=4$ qubits. Fig.~\ref{fig:fidelity_training} plots the fidelity for each of the six states in the denoising path, while Fig.~\ref{fig:loss_training} shows the corresponding convergence of the global path-constrained loss over training steps.

The simultaneous analysis of the intermediate state fidelity in Fig.~\ref{fig:fidelity_training} and the global loss in Fig.~\ref{fig:loss_training} reveals that the optimization successfully converges to a global objective, rather than achieving a collection of local optima for each intermediate state.

This is most evident in the complex, non-monotonic behavior of the intermediate fidelities in Fig.~\ref{fig:fidelity_training}. Across all system sizes (a-d), we observe that the optimization does not improve all intermediate states uniformly. Specifically, one intermediate state (red curve, $P3$) consistently shows a decrease in fidelity from its initial value, while another state (green curve, $P2$) remains at a relatively low, stagnant fidelity throughout the training. In stark contrast, the fidelity of the other states, including the final target state, are successfully optimized to high values. This behavior indicates that the optimization is finding a globally optimal trajectory by enforcing trade-offs; it appears to sacrifice fidelity at certain intermediate steps to achieve a greater fidelity gain for the final target state. This result is consistent with the trajectory analysis in Figs.~\ref{fig:bloch_evo_6} and \ref{fig:bloch_evo_12}, which showed that the learned backward path is not a perfect time-reversal of the diffusion, but rather a distinct, physically valid path that satisfies the global optimization. These complex trade-offs are established very early in the training process. Both the fidelity curves and the loss curves exhibit their most dramatic changes within the first 100 time steps. Despite the complex internal dynamics and sacrifices at certain nodes, the smooth and rapid convergence of the overall loss function (in Fig.~\ref{fig:loss_training}) confirms that the holistic optimization is stable and effective. This intricate balance between local path constraints and the global objective of target state generation highlights the non-trivial nature of the learned solution and warrants further theoretical exploration.
\end{document}